\documentclass[12pt,letterpaper]{article}

\usepackage{epsfig,amsmath,amssymb,latexsym,graphicx,setspace,fullpage}
\def\bib{\vskip12pt\par\noindent\hangindent=0.5 true cm\hangafter=1}

\begin{document}

\renewcommand{\thefootnote}{\fnsymbol{footnote}}
\begin{center}
{\LARGE Variational approximation for \\mixtures of linear mixed models\\}
\end{center}

\begin{center}
{\large Siew Li Tan and David J. Nott \footnote{  
Siew Li Tan is PhD student, Department of Statistics and Applied Probability, National University
of Singapore, Singapore 117546 (email g0900760@nus.edu.sg). David J. Nott is Associate Professor, Department of Statistics and Applied Probability, National University of Singapore, Singapore 117546. (email  standj@nus.edu.sg).}} \\
\end{center}

\begin{abstract}
Mixtures of linear mixed models (MLMMs) are useful for clustering grouped data and can be estimated by likelihood maximization through the EM algorithm. The conventional approach to determining a suitable number of components is to compare different mixture models using penalized log-likelihood criteria such as BIC. We propose fitting MLMMs with variational methods which can perform parameter estimation and model selection simultaneously. A variational approximation is described where the variational lower bound and parameter updates are in closed form, allowing fast evaluation. A new variational greedy algorithm is developed for model selection and learning of the mixture components. This approach allows an automatic initialization of the algorithm and returns a plausible number of mixture components automatically. In cases of weak identifiability of certain model parameters, we use hierarchical centering to reparametrize the model and show empirically that there is a gain in efficiency by variational algorithms similar to that in MCMC algorithms. Related to this, we prove that the approximate rate of convergence of variational algorithms by Gaussian approximation is equal to that of the corresponding Gibbs sampler which suggests that reparametrizations can lead to improved convergence in variational algorithms as well.
\end{abstract}


\section{Introduction}
Mixtures of linear mixed models (MLMMs) are useful for clustering grouped data in applications such as clustering of gene expression profiles (Celeux {\it et al.}, 2005, and Ng {\it et al.}, 2006) and electrical load series (Coke and Tsao, 2010). We consider MLMMs where the response distribution is a normal mixture with the mixture weights varying as a function of the covariates. Our model includes cluster-specific random effects so that observations from the same cluster are correlated. We propose fitting MLMMs with variational methods which can perform parameter estimation and model selection 
simultaneously. Our article makes four contributions. First, fast variational
methods are developed for MLMMs and a variational lower bound is obtained in closed 
form. Second, a new variational greedy algorithm is developed for model 
selection and learning of the mixture components. This approach handles  
algorithm initialization and returns a plausible number of mixture components automatically. Third, we show empirically that there is a gain in efficiency by variational 
algorithms through the use of hierarchical centering reparametrization 
similar to that in Markov chain Monte Carlo (MCMC) algorithms. Fourth, we prove that the approximate 
rate of convergence of the variational algorithm by Gaussian approximation is equal
to that of the corresponding Gibbs sampler which suggests that reparametrizations can give improved convergence in variational algorithms just as in MCMC algorithms.

In microarray analysis, clustering of gene expression profiles is a valuable exploratory tool 
for the identification of potentially meaningful relationships between genes.
In the model-based cluster analysis context, Luan and Li (2003) studied
the clustering of genes based on time course gene expression profiles 
in the mixture model framework using a mixed-effects model with B-splines.
Celeux {\it et al.} (2005) proposed using MLMMs to account for data variability in repeated measurements. Both of these approaches require the independence assumption for genes.
In contrast, Ng {\it et al.} (2006) considered MLMMs which allow genes within 
a cluster to be correlated as the independence assumption may not hold for 
all pairs of genes (McLachlan {\it et al.}, 2004). Booth {\it et al.} (2008) considered
a multilevel linear mixed model (LMM) which includes cluster-specific random effects 
and proposed a stochastic search algorithm for finding partitions of the data with 
high posterior probability through maximization of an objective function. 
For the clustering of electrical load series, Coke and Tsao (2010) developed random effects mixture models using a hierarchical representation and used an antedependence model for the non-stationary random effects. 

MLMMs can be estimated by likelihood maximization through the EM algorithm (Dempster {\it et al.}, 1977) and this method was used in Luan and Li (2003), Celeux {\it et al.} (2005) and Coke and Tsao (2010). Ng {\it et al.} (2006) developed a program called EMMIX-WIRE (EM-based MIXture analysis WIth Random Effects) for clustering correlated and replicated data. The optimal number of components was determined by comparing different mixture models using the BIC (Bayesian information criterion) of Schwarz (1978) in these articles.  The EM algorithm can be sensitive to initialization and is commonly run from multiple starting values to avoid convergence to local optima. Scharl {\it et al.} (2010) studied the performance of different initialization strategies for mixtures of regression models. In the context of Gaussian mixture models, Biernacki {\it et al.} (2003) compared simple initialization strategies and Verbeek {\it et al.} (2003) discussed a greedy approach to the learning of Gaussian mixtures which resolves the sensitivity to initialization and is useful in finding the optimal number of components.

We propose fitting MLMMs with variational methods using a greedy algorithm. The MLMM we consider is a simple generalization of that proposed by Ng {\it et al.} (2006) where units within each cluster may be correlated. A variational approximation for this model is 
described where the variational lower bound and parameter updates are in closed form, allowing fast evaluation. Ormerod and Wand (2010) illustrated the use of variational methods to fit a Gaussian LMM and 
Armagan and Dunson (2011) used variational methods to obtain sparse approximate 
Bayes inference in the analysis of large longitudinal data sets using LMMs. 
Ormerod and Wand (2012) recently introduced an approach called Gaussian variational approximation for fitting 
generalized LMMs where the distributions of random effects vectors are approximated by Gaussian distributions. The variational algorithm suffers from problems of local optima and initialization strategies for the EM algorithm can often be adapted for use with the variational algorithm. A common strategy is to run the variational algorithm starting with random initialization from 
multiple starting points (Bishop and Svens\'{e}n, 2003). Nott {\it et al.} (2011) used a short runs 
strategy similar to that recommended by Biernacki {\it et al.} (2003) where the 
variational algorithm is stopped prematurely and only the short run with the highest 
attained value of the variational lower bound is followed to convergence. 

A key advantage of variational methods is the potential for simultaneous parameter
estimation and model selection and a number of such methods have been developed
for the fitting of Gaussian mixtures. Ueda and Ghahramani (2002) proposed 
a variational Bayesian (VB) split and merge EM procedure to 
optimize an objective function that allows simultaneous estimation of the 
parameters and the number of components while avoiding local optima. 
They applied this method to a Gaussian mixture and a mixture of experts regression where both input and output are treated as random variables. Wu {\it et al.} (2012) developed a split and eliminate VB algorithm which attempts to split all poorly fitted components at the 
same time and made use of the component-elimination property 
associated with variational approximation so that no merge moves are required. 
This component-elimination property was noted previously by Attias (1999) and Corduneanu and Bishop (2001).
McGrory and Titterington (2007) described a variational optimization 
technique where the algorithm is initialised with a large number of components and 
mixture components whose weightings become sufficiently small are dropped out as the 
optimization proceeds. Constantinopoulos and Likas (2007) observed that in this approach, the number of components in the resulting mixture can be sensitive to the prior on the precision matrix. They proposed an incremental approach where components are added to the
mixture following a splitting test where a different local precision prior is specified after
taking into account characteristics of the precision matrix of the component being tested.

For the examples in this paper, we have attempted the component deletion approach of McGrory and Titterington (2007) (results not shown). We observed that this method is more effective when the number of components required is not too large as initializing the mixture with a large number of components can be computationally expensive especially for large data sets. The choice of the initial number of mixture
components can have an impact on the resulting number of components and it may not be 
easy to determine a suitable initial number. This approach remains sensitive to initialization and methods such as running the variational algorithm from multiple starting 
points are necessary to avoid local optima.

We develop a new variational greedy algorithm (VGA) for the learning of MLMMs. 
This greedy approach is not limited to MLMMs and may be extended
to fit other models using variational methods. No additional derivations are required once the basic variational algorithm is available. Starting with one component, the
VGA adds new components to the mixture after searching for the 
optimal way to split components in the current mixture. 
This approach handles algorithm initialization automatically and returns a plausible value for the 
number of mixture components. While this bottom-up approach resolves the difficulty of estimating the upper bound of the number of mixture components, it can become time-consuming when the number of components is large, since a larger number of components have to be tested to find the optimal way of splitting each one. Some measures are introduced to keep the search time short and the component elimination property of variational approximation is used to sieve out components which resist splitting.

In situations where there is weak identification of certain model parameters and the variational algorithm converges very slowly, we apply hierarchical centering (Gelfand {\it et al.}, 1995) to reparametrize the MLMM. Hierarchical centering has been applied successfully in MCMC algorithms to obtain improved convergence (Chen {\it et al.}, 2000) and we show empirically, that there is a similar gain in efficiency in variational algorithms. We consider a case of partial centering, a second case of full centering and derive the corresponding variational algorithms.
Related to this, we show that the approximate rate of convergence of the variational algorithm by 
Gaussian approximation is equal to that of the corresponding Gibbs sampler. Sahu and Roberts (1999) showed that the approximate rate of convergence of the Gibbs sampler 
by Gaussian approximation is equal to that of the corresponding EM-type algorithm and hence 
improvement strategies for one algorithm can be used for the other. 
As reparametrizations using hierarchical centering can lead to improved convergence in the Gibbs sampler,
this result suggests that the rate of convergence of variational algorithms may be improved through reparametrizations. Papaspiliopoulos {\it et al.} (2007) describe centering and non-centering methodology as complementary techniques for use in parametrization of hierarchical models to construct effective MCMC algorithms. 

In Section 2, we introduce MLMMs. Section 3 describes fast variational approximation methods for MLMMs and Section 4 reparametrization of MLMMs through hierarchical centering. Section 5 describes the variational greedy algorithm and Section 6 contains theoretical results on the rate of convergence of variational algorithms by Gaussian approximation. Section 7 considers examples involving real and simulated data and Section 8 concludes.

\section{Mixtures of linear mixed models}
The MLMM we consider is a generalization of that proposed by Ng {\it et al.} (2006), where units from the same cluster share cluster-specific random effects and are hence correlated. Unlike Ng {\it et al.} (2006), our model can fit data where the number of observations on each unit are not equal and we allow the mixture weights to vary with covariates between clusters. Suppose we observe $n$ multivariate reponses $y_i=(y_{i1},...,y_{in_i})^T$, $i=1,...,n$ and $N=\sum_{i=1}^n n_i$.
Let the number of mixture components be $k$ and $z_i$, $i=1,...,n$ be latent variables indicating which mixture component the $i$th cluster corresponds to, $z_i \in \{1,...,k\}$. Conditional on $z_i=j$,
\begin{eqnarray}
y_i = X_i \beta_j + W_i a_i + V_i b_j + \epsilon_i \label{MLMM}
\end{eqnarray}
where $X_i$, $W_i$ and $V_i$ are design matrices of dimensions $n_i \times p$, $n_i \times s_1$ and 
$n_i \times s_2$ respectively, $\beta_j$, $j=1,...,k,$ are $p \times 1$ vectors of fixed effects, $a_i$,
$i=1,...,n,$ are $s_1 \times 1$ vectors of random effects, $b_j$, $j=1,...,k,$ are $s_2 \times 1$ vectors of random effects and $\epsilon_i$, $i=1,...,n,$ are vectors of random errors. We assume that the random effects $a_i$, $i=1,...,n$, $b_j$, $j=1,...,k,$ and the error vectors $\epsilon_i$, $i=1,...,n,$ are mutually independent. The fixed effects, the distribution of the random effects 
and the distribution of the error terms are all mixture component specific.
The random effects distribution for $a_i$ and $b_j$ are $N( 0,\sigma_{a_j}^2 I_{s_1})$ 
and $N( 0,\sigma_{b_j}^2 I_{s_2})$ respectively. 
The error vector $\epsilon_i$ is distributed as $N \left( 0,\Sigma_{ij} \right)$ where $\Sigma_{ij}=\mbox{blockdiag}(\sigma_{j1}^2I_{\kappa_{i1}},...,\sigma_{jg}^2I_{\kappa_{ig}})$,
a block diagonal with the $l$th block equal to $\sigma_{jl}^2I_{\kappa_{il}}$. Here $g$ is constant for all $i$ and $\sum_{l=1}^g {\kappa_{il}}=n_i$ for each $i=1,...,n$. In microarray experiments for instance, this specification provides increased flexibility as the error variance of each mixture component is allowed to vary between different experiments, say, by setting $g$ to be the total number of experiments.
We assume that
\begin{eqnarray}
P(z_i=j)=p_{ij}=\frac{\exp(u_i^T \delta_j)}{\sum_l \exp(u_i^T \delta_l)} \label{gating}
\end{eqnarray} 
where $u_i=(u_{i1},...,u_{id})^T$ is a vector of covariates and 
$\delta_j=(\delta_{j1},...,\delta_{jd})^T$ are vectors of unknown parameters
$j=2,...,k$. We set $\delta_1=0$ for identifiability. These mixing coefficients,
which are functions of the covariates, are known as gating functions in the 
mixture of experts terminology (Jacobs {\it et al.}, 1991).  
This model for the mixture component indicators allows the mixture weights
to vary with covariates across clusters. For Bayesian inference on unknown
parameters we assume the following priors. $\sigma_{a_j}^2 \sim IG(\alpha_{a_j},\lambda_{a_j})$,
$j=1,...,k,$ where $IG(\alpha,\lambda)$ denotes the inverse gamma distribution with shape parameter 
$\alpha$ and scale parameter $\lambda$, $\sigma_{b_j}^2 \sim IG(\alpha_{b_j},\lambda_{b_j})$, $j=1,...,k$, 
$\sigma_{jl}^2 \sim IG(\alpha_{jl},\lambda_{jl})$, $j=1,...,k$, $l=1,...,g$,
$\delta=(\delta_2^T,...,\delta_k^T)^T\sim N(0,\Sigma_\delta)$ and 
$\beta_j\sim N(0,\Sigma_{\beta j})$. Here $\alpha_{a_j}$, $\lambda_{a_j}$, $\alpha_{b_j}$, $\lambda_{b_j}$,
$\alpha_{jl}$, $\lambda_{jl}$, $\Sigma_{\delta}$ and $\Sigma_{\beta_j}$, $j=1,...,k$, $l=1,...,g$,
are hyperparameters considered known.

\section{Variational approximation}
Variational methods originated in statistical physics and research into these approaches is currently very active in both statistics and machine learning. Until recently, variational approximation 
methods have mostly been developed in the machine learning community (Jordan {\it et al.}, 1999, 
Winn and Bishop, 2005). See Ormerod and Wand (2010) for an explanation of variational approximation methods and the introduction for further references on application of variational methods to mixture models specifically. We consider a variational approximation to the joint posterior distribution of 
all the parameters $\theta$ of the form $q(\theta|\lambda)$ where $\lambda$ 
is the set of variational parameters to be chosen.  Here a parametric form is chosen 
for $q(\theta|\lambda)$ and we attempt to make $q(\theta|\lambda)$ 
a good approximation to $p(\theta|y)$ by minimizing the Kullback-Leibler (KL) divergence 
between $q(\theta|\lambda)$ and $p(\theta|y)$, i.e.,
$$\int \log \frac {q(\theta|\lambda)}{p(\theta|y)} q(\theta|\lambda)\;d\theta 
= \int \log \frac {q(\theta|\lambda)}{p(\theta)p(y|\theta)} q(\theta|\lambda)\;d\theta + \log p(y)$$ 
where $p(y) = \int p(y|\theta)p(\theta)\;d\theta$ is the marginal likelihood. 
As the KL divergence is positive,
$$\log p(y)\geq \int \log \frac {p(\theta)p(y|\theta)}{q(\theta|\lambda)} q(\theta|\lambda)\;d\theta $$
which gives a lower bound on the log marginal likelihood, and maximization of this lower bound 
is equivalent to minimisation of the KL divergence between the posterior 
distribution and variational approximation. This lower bound is sometimes used as an approximation to the log marginal likelihood for Bayesian model selection purposes.

Write $\beta=(\beta_1^T,...,\beta_k^T)^T$, $a=(a_1^T,...,a_n^T)^T$, $b=(b_1^T,...,b_k^T)^T$, 
$\sigma_a^2=(\sigma_{a1}^2,...,\sigma_{ak}^2)^T$, $\sigma_b^2=(\sigma_{b1}^2,...,\sigma_{bk}^2)^T$,
$\sigma_j^2=(\sigma_{j1}^2,...,\sigma_{jg}^2)^T$, $j=1,...,k$, $\sigma^2=({\sigma_1^2}^T,...,{\sigma_k^2}^T)^T$,
$\delta=(\delta_2^T,...,\delta_k^T)^T$ and $z=(z_1,...,z_n)^T$ 
so that $\theta=(\beta^T,a^T,b^T,{\sigma_a^2}^T,{\sigma_b^2}^T,{\sigma^2}^T,\delta^T,z^T)^T$. For convenience
we write $q(\theta|\lambda)$ as $q(\theta)$, suppressing dependence 
on $\lambda$ and consider a variational approximation of the form
$q(\theta)=q(\beta)q(a)q(b)q(\sigma_a^2)q(\sigma_b^2)q(\sigma^2)q(\delta)q(z)$, where
\begin{gather*}
q(\beta)=\prod_{j=1}^k q(\beta_j),\;q(a)=\prod_{i=1}^n q(a_i),\;q(b)=\prod_{j=1}^k q(b_j),\;q(z)=\prod_{i=1}^n q(z_i), \\  
q(\sigma_a^2)=\prod_{j=1}^k q(\sigma_{a_j}^2),\; q(\sigma_b^2)=\prod_{j=1}^k q(\sigma_{b_j}^2),\; 
q(\sigma^2)=\prod_{j=1}^k \prod_{l=1}^g q(\sigma_{jl}^2), 
\end{gather*}
and $q(\beta_j)$ is $N(\mu_{\beta_j}^q,\Sigma_{\beta_j}^q)$, $j=1,...,k$, 
$q(a_i)$ is $N(\mu_{a_i}^q,\Sigma_{a_i}^q)$, $i=1,...,n,$
$q(b_j)$ is $N(\mu_{b_j}^q,\Sigma_{b_j}^q)$, $j=1,...,k$, 
$q(\sigma_{a_j}^2)$ is $IG(\alpha_{a_j}^q,\lambda_{a_j}^q)$, $j=1,...,k$,
$q(\sigma_{b_j}^2)$ is $IG(\alpha_{b_j}^q,\lambda_{b_j}^q)$, $j=1,...,k$,
$q(\sigma_{jl}^2)$ is $IG(\alpha_{jl}^q,\lambda_{jl}^q)$, for $j=1,...,k$, $l=1,...,g$,
$q(\delta)$ is a delta function placing a point mass of 1 on $\mu_\delta^q$, and 
$q(z_i=j)=q_{ij}$ where $\sum_{j=1}^k q_{ij}=1$, $i=1,...,n$. We are assuming in the 
variational posterior that parameters for different expert components are independent and 
independent of all other parameters. 

For a variational posterior restricted to be of the factorized form $q(\theta)=\prod_{i=1}^m q(\theta_i)$, the optimal $q(\theta_i)$ minimizing the KL divergence is given by $q(\theta_i) \propto \exp\{  E_{-\theta_i}\log p(y,\theta)\}$ (see, for example, Ormerod and Wand, 2010). In our case, the specific distributional forms for the variational posterior densities, such as the assumption of a degenerate point mass variational posterior for $\delta$, have been chosen to make computation of the lower bound tractable even though they might not be optimal. It is also possible to consider the fixed effects $\beta$, and the random effects $a$ and $b$ as a single block and replace $q(\beta)q(a)q(b)$ by $q(\beta,a,b)$ as in Ormerod and Wand (2010). This results in a less restricted factorization with dependence structure between $\beta$, $a$ and $b$ preserved and a higher lower bound can be achieved. However, this will involve dealing with high dimensional sparse covariance matrices which creates a greater computational burden although it is possible to use matrix inversion results for the blocked matrices to attain better computational efficiency (referee's suggestion). We have decided to use a factorized form for faster computation and better scalability to larger data sets (see Armagan and Dunson, 2011). The independence and distributional assumptions made in variational approximations may not be realistic and it has been shown in the context of Gaussian mixture models that VB, which assumes a factorized posterior, has a tendency to underestimate the posterior variance (Wang and Titterington, 2005). However, variational approximation can often lead to good point estimates, reasonable estimates of marginal posterior distributions and excellent predictive inferences. Blei and Jordan (2006) showed that predictive distributions based on variational approximations to the posterior were very similar to that of MCMC for Dirichlet process mixture models. Braun and McAuliffe (2010) reported similar findings in large-scale models of discrete choice although they observed that the variational posterior is more concentrated around the mode than the MCMC posterior, a familiar underdispersion effect noted above. Similar independence assumptions have been made in the case of the LMM by Armagan and Dunson (2011).

Now, we want to maximise the variational lower bound $L=\int\log \frac{p(\theta)p(y|\theta)}{q(\theta)}q(\theta)\;d\theta$ 
with respect to the parameters $\lambda$ in our variational posterior approximation. 
The lower bound $L$ can be computed in closed form, and is given by (details in supplementary materials)
\begin{align}
& \frac{1}{2}\sum_{j=1}^k \left\{ \log |\Sigma_{\beta_j}^{-1} \Sigma_{\beta_j}^q|
  - \mbox{tr} (\Sigma_{\beta_j}^{-1} \Sigma_{\beta_j}^q) 
  - {\mu_{\beta_j}^q}^T \Sigma_{\beta_j}^{-1} \mu_{\beta_j}^q  + \log |\Sigma_{b_j}^q| 
  - \frac{\alpha_{b_j}^q}{\lambda_{b_j}^q} 
    \left( {\mu_{b_j}^q}^T \mu_{b_j}^q + \mbox{tr} (\Sigma_{b_j}^q) \right)\right\}\nonumber  \\
& + \sum_{j=1}^k \left\{ \alpha_{b_j} \log \frac{\lambda_{b_j}}{\lambda_{b_j}^q} 
  + \log \frac{\Gamma(\alpha_{b_j}^q)}{\Gamma(\alpha_{b_j})}- \frac{\lambda_{b_j} \alpha_{b_j}^q}{\lambda_{b_j}^q}- \frac{s_2}{2}\log(\lambda_{b_j}^q) 
  + \alpha_{b_j}^q+\alpha_{a_j} \log \frac{\lambda_{a_j}}{\lambda_{a_j}^q} 
  + \log \frac{\Gamma(\alpha_{a_j}^q)}{\Gamma(\alpha_{a_j})} \right. \nonumber\\
& \left. - \frac{s_1\sum_{i=1}^n q_{ij}}{2}(\psi(\alpha_{a_j}^q)-\log(\lambda_{a_j}^q))   
  +\psi(\alpha_{a_j}^q)(\alpha_{a_j}-\alpha_{a_j}^q)  
  - \frac{\lambda_{a_j} \alpha_{a_j}^q}{\lambda_{a_j}^q} +\alpha_{a_j}^q \right\} 
  + \frac{1}{2}\sum_{i=1}^n \log |\Sigma_{a_i}^q|\nonumber \\
& + \sum_{j=1}^k \sum_{l=1}^g \left\{ \alpha_{jl} \log \frac{\lambda_{jl}}{\lambda_{jl}^q}
  + \log \frac{\Gamma(\alpha_{jl}^q)}{\Gamma(\alpha_{jl})} 
  + \frac{\sum_{i=1}^n\kappa_{il}q_{ij}}{2} 
    \left(  \psi(\alpha_{jl}^q)-\log(\lambda_{jl}^q) \right)   
  + \psi(\alpha_{jl}^q)(\alpha_{jl}-\alpha_{jl}^q)\right. \nonumber\\
& \left. - \frac{\lambda_{jl} \alpha_{jl}^q}{\lambda_{jl}^q} + \alpha_{jl}^q \right\}
  -\sum_{i=1}^n \sum_{j=1}^k \frac{q_{ij}}{2} \left\{\xi_{ij}^T {\Sigma_{ij}^q}^{-1} \xi_{ij}+
    \mbox{tr} ({\Sigma_{ij}^q}^{-1} \Lambda_{ij})+\frac{\alpha_{a_j}^q}{\lambda_{a_j}^q} 
    \left( {\mu_{a_i}^q}^T \mu_{a_i}^q + \mbox{tr} (\Sigma_{a_i}^q) \right) \right\} \nonumber\\
&  +\sum_{i=1}^n \sum_{j=1}^k q_{ij} \log \frac{p_{ij}}{q_{ij}}+\log p(\mu_\delta^q)
  +\frac{k(p+s_2)+ns_1-N\log (2\pi)}{2}    \label{lb_alg1}
\end{align}
where $\Gamma(\cdot)$ and $\psi(\cdot)$ denote the gamma and digamma functions respectively, $p_{ij}$ is evaluated by setting $\delta=\mu_{\delta}^q$, $p(\mu_\delta^q)$ denotes the prior distribution for $\delta$ evaluated at $\mu_\delta^q$,  $\xi_{ij}=y_i-X_i\mu_{\beta_j}^q-W_i\mu_{a_i}^q-V_i \mu_{b_j}^q$,
${\Sigma_{ij}^q}^{-1} = \mbox{blockdiag} \left(\frac{\alpha_{j1}^q}{\lambda_{j1}^q}I_{\kappa_{i1}}, ...,
\frac{\alpha_{jg}^q}{\lambda_{jg}^q}I_{\kappa_{ig}} \right)$ and $\Lambda_{ij}=X_i\Sigma_{\beta_j}^qX_i^T+W_i\Sigma_{a_i}^q W_i^T+V_i \Sigma_{b_j}^q V_i^T$. The variational parameters to be optimized consist of $\mu_{\beta_j}^q$, $\Sigma_{\beta_j}^q$, $\mu_{b_j}^q$, $\Sigma_{b_j}^q$, $\alpha_{a_j}^q$, $\lambda_{a_j}^q$, $\alpha_{b_j}^q$, $\lambda_{b_j}^q$, for $j=1,...,k$, $\mu_{a_i}^q$, $\Sigma_{a_i}^q$, for $i=1,...,n$, $\alpha_{jl}^q$, $\lambda_{jl}^q$, for $j=1,...,k$, $l=1,...,g$, $q_{ij}$ for $i=1,...,n$, $j=1,...,k,$ and $\mu_\delta^q$. We optimize the lower bound with respect to each of these sets of parameters with the others held fixed in a gradient ascent algorithm. All updates except for $\mu_\delta^q$ are available in closed form and can be derived using vector differential calculus (see Wand, 2002). 

\vspace{3mm}
\noindent {\it \textbf{Algorithm 1}:}\\
Initialize: $q_{ij}$ for $i=1,...,n$, $j=1,...,k$,
            $\frac{\alpha_{jl}^q}{\lambda_{jl}^q}$ for $j=1,...,k$, $l=1,...,g$,
            $\mu_{a_i}^q$ for $i=1,...,n$, $\mu_{b_j}^q$,
            $\frac{\alpha_{a_j}^q}{\lambda_{a_j}^q}$ and $\frac{\alpha_{b_j}^q}{\lambda_{b_j}^q}$
            for $j=1,...,k$.         
Do until the change in the lower bound between iterations is less than a tolerance:
\begin{enumerate}
\item $\Sigma_{\beta j}^q \leftarrow (\Sigma_{\beta j}^{-1}+\sum_{i=1}^n q_{ij} X_i^T {\Sigma_{ij}^q}^{-1} X_i)^{-1}$.
\item $\mu_{\beta j}^q \leftarrow \Sigma_{\beta j}^q \sum_{i=1}^n q_{ij} X_i^T {\Sigma_{ij}^q}^{-1} (y_i-W_i\mu_{a_i}^q-V_i\mu_{b_j}^q).$
\item $\Sigma_{a_i}^q \leftarrow \left(\sum_{j=1}^k q_{ij} \frac{\alpha_{a_j}^q}{\lambda_{a_j}^q} I_{s_1} 
		  + W_i^T (\sum_{j=1}^k q_{ij} {\Sigma_{ij}^q}^{-1}) W_i \right)^{-1}. $
\item $ \mu_{a_i}^q \leftarrow \Sigma_{a_i}^q \sum_{j=1}^k q_{ij} W_i^T {\Sigma_{ij}^q}^{-1} (y_i-X_i \mu_{\beta_j}^q-V_i \mu_{b_j}^q). $
\item $ \Sigma_{b_j}^q \leftarrow \left( \frac{\alpha_{b_j}^q}{\lambda_{b_j}^q}I_{s_2} 
      + \sum_{i=1}^n q_{ij} V_i^T {\Sigma_{ij}^q}^{-1} V_i \right)^{-1}. $
\item $ \mu_{b_j}^q \leftarrow \Sigma_{b_j}^q \sum_{i=1}^n q_{ij} V_i^T {\Sigma_{ij}^q}^{-1} (y_i-X_i \mu_{\beta_j}^q-W_i\mu_{a_i}^q).$
\item $ \alpha_{a_j}^q \leftarrow \alpha_{a_j} + \frac{s_1}{2}\sum_{i=1}^n q_{ij}.$
\item $ \lambda_{a_j}^q \leftarrow \lambda_{a_j} + \frac{1}{2} \sum_{i=1}^n q_{ij} 
        \{ {\mu_{a_i}^q}^T\mu_{a_i}^q +\mbox{tr}(\Sigma_{a_i}^q) \}.$
\item $ \alpha_{b_j}^q \leftarrow \alpha_{b_j} + \frac{s_2}{2}.$
\item $ \lambda_{b_j}^q \leftarrow \lambda_{b_j} + \frac{1}{2} \{ {\mu_{b_j}^q}^T\mu_{b_j}^q
      +\mbox{tr}(\Sigma_{b_j}^q)\}.$
\item $ \alpha_{jl}^q \leftarrow \alpha_{jl} + \frac{1}{2}\sum_{i=1}^n q_{ij}\kappa_{il}.$
\item $ \lambda_{jl}^q \leftarrow \lambda_{jl} + \frac{1}{2} \sum_{i=1}^n q_{ij} \{
      (\xi_{ij})_{\kappa_{il}}^T (\xi_{ij})_{\kappa_{il}}+ 
     \mbox{tr} (\Lambda_{ij})_{\kappa_{il}} \}$ \\
where $((\xi_{ij})_{\kappa_{i1}},...,(\xi_{ij})_{\kappa_{ig}})$ is the 
partition of $\xi_{ij}$ corresponding to the $(\kappa_{i1},...,\kappa_{ig})$
and $(\Lambda_{ij})_{\kappa_{il}}$ is the diagonal block of $\Lambda_{ij}$ with 
rows and columns corresponding to the position of $\kappa_{il}$ within 
$(\kappa_{i1},...,\kappa_{ig})$.
\item Set $\mu_{\delta}^q$ to be the conditional mode of the lower bound fixing 
other variational parameters at their current values. As a function of $\mu_{\delta}^q$,
the lower bound is the log posterior for a Bayesian multinomial regression with the $i$th response being  $(q_{i1},...,q_{ik})^T$ and a normal prior on $\mu_\delta^q$. The usual iteratively weighted least squares algorithm (or other numerical optimization algorithm) can be used for finding the mode.  
\item $q_{ij} \leftarrow \frac{p_{ij} \exp(c_{ij})}{\sum_{l=1}^k p_{il}\exp(c_{il})},$ where
$c_{ij}= \frac{s_1}{2} \{ \psi(\alpha_{a_j}^q)-\log (\lambda_{a_j}^q)\}
-\frac{\alpha_{a_j}^q}{2\lambda_{a_j}^q}\{ {\mu_{a_i}^q}^T\mu_{a_i}^q+\mbox{tr}(\Sigma_{a_i}^q) \} 
+\frac{1}{2} \sum_{l=1}^g \kappa_{il} \{ \psi(\alpha_{jl}^q)-\log (\lambda_{jl}^q)\}  
- \frac{1}{2} \{ \mbox{tr} ({\Sigma_{ij}^q}^{-1} \Lambda_{ij}) + \xi_{ij}^T {\Sigma_{ij}^q}^{-1}
  \xi_{ij} \}$.
\end{enumerate} 
In the examples, when Algorithm 1 is used in conjunction with the VGA described in Section 5
to fit a 1-component mixture, for $j=1$, we set
$\frac{\alpha_{a_j}^q}{\lambda_{a_j}^q}=\frac{\alpha_{b_j}^q}{\lambda_{b_j}^q}=1$,
$\frac{\alpha_{jl}^q}{\lambda_{jl}^q}=1$ for $l=1,...,g$,
$\mu_{b_j}^q=0$, $\mu_{a_i}^q=0$ for $i=1,...,n$, and
$q_{ij}=1$ for $i=1,...,n$ for initialization.

The variational posterior for $\delta$ has been assumed to be a degenerate point mass to make computation of the lower bound tractable. However, at convergence, we relax the form of $q(\delta)$ to be a normal distribution. Suppose $q(\delta)$ is not subjected to any distributional restriction, the optimal choice for this term is given by 
\begin{align}
q(\delta) \propto \exp \left \{ \sum_{i=1}^n \sum_{j=1}^k q_{ij} \log p_{ij} -\frac{1}{2}\delta^T \Sigma_\delta^{-1} \delta  \right\} \;\;\;\text{where}\;\;\; p_{ij} =\frac{\exp(u_i^T \delta_j)}{\sum_l \exp(u_i^T \delta_l)}. \label{mulpost}
\end{align}
If $\mu_\delta^q$ is close to the mode, we can get a normal approximation to $q(\delta)$ by setting $\mu_\delta^q$ as the mean and the covariance matrix $\Sigma_\delta^q$ as the negative inverse Hessian of the log of (\ref{mulpost}) which is the Bayesian multinomial log posterior considered in step 13 of Algorithm 1. Waterhouse {\it et al.} (1996) outlined a similar idea which they used at every step of their iterative algorithm. We recommend first using a delta function approximation in the VGA and doing a one-step approximation after the algorithm has converged (see Nott {\it et al.}, 2011). Using the normal approximation $N(\mu_\delta^q,\Sigma_\delta^q)$ as the variational posterior for $q(\delta)$, the variational lower bound is the same as in (\ref{lb_alg1}) except that $\sum_{i=1}^n \sum_{j=1}^k q_{ij} \log p_{ij} + \log p(\mu_\delta^q)$ is replaced with 
$$ \sum_{i=1}^n \sum_{j=1}^k q_{ij} E_q \left(\log \frac{\exp(u_i^T \delta_j)}{\sum_l \exp(u_i^T \delta_l)}  \right)+\frac{1}{2} \log |\Sigma_{\delta}^{-1}\Sigma_\delta^q|-\frac{1}{2}{\mu_\delta^q}^T\Sigma_\delta^{-1}\mu_\delta^q-\frac{1}{2}\text{tr}(\Sigma_\delta^{-1}\Sigma_\delta^q)+\frac{d(k-1)}{2}.$$
The expectation of the first term, $E_q \left(\log \left \{ \frac{\exp(u_i^T \delta_j)}{\sum_l \exp(u_i^T \delta_l)} \right \} \right)$, is not available in closed form and we replace it with $\log \left \{ \frac{\exp(u_i^T \mu_{\delta_j}^q)}{\sum_l \exp(u_i^T \mu_{\delta_l}^q)} \right \} $
where $\mu_{\delta_j}^q$ is the subvector of $\mu_\delta^q$ corresponding to $\delta_j$, $j=2,...,k$, to obtain an approximation to $\log p(y)$. This approximation to the log marginal likelihood is later used in the VGA as a model selection criterion. 

\section{Hierarchical Centering}
Gelfand {\it et al.} (1995) discussed how reparametrizations of normal LMMs using hierarchical centering can improve convergence in MCMC algorithms. In our later examples we encounter situations where there is weak identification of certain model parameters and Algorithm 1 converges very slowly. We apply hierarchical centering and show empirically that there is a gain in efficiency in variational algorithms through hierarchical centering, similar to that in MCMC algorithms. In Section 6 we give some theoretical support for this observation.

We consider a case of partially centered parametrization in which $X_i=W_i$ and a second case of fully centered parametrization in which $X_i=W_i=V_i$ in (\ref{MLMM}). In the first case, we introduce $\eta_i=\beta_j+a_i$ conditional on $z_i=j$ so that (\ref{MLMM}) is reparametrized as
\begin{eqnarray}
y_i = X_i \eta_i + V_i b_j + \epsilon_i \nonumber
\end{eqnarray}
and $\eta_i$ is `centered' about $\beta_j$, with $\eta_i\sim N(\beta_j,\sigma_{a_j}^2 I_p)$.
Writing $\eta=(\eta_1^T,...,\eta_n^T)^T$, the set of unknown parameters is now $\theta=(\beta^T,\eta^T,b^T,{\sigma_a^2}^T,{\sigma_b^2}^T,{\sigma^2}^T,\delta^T,z^T)^T$. 
We replace $q(a)$ in the variational approximation with $q(\eta)=\prod_{i=1}^n q(\eta_i)$,
where $q(\eta_i)$ is $N(\mu_{\eta_i}^q,\Sigma_{\eta_i}^q)$, for $i=1,...,n$. In the second case of full centering, we introduce $\rho_i=\nu_j+a_i$ and $\nu_{j}=\beta_j+b_j$, conditional on $z_i=j$ so that (\ref{MLMM}) is reparametrized as
\begin{eqnarray}
y_i = X_i \rho_{i} + \epsilon_i, \nonumber
\end{eqnarray}
with $\rho_i$ `centered' about $\nu_j$ and $\nu_j$ `centered' about $\beta_j$. We have $\rho_i\sim N(\nu_j,\sigma_{a_j}^2 I_p)$ and $\nu_j\sim N(\beta_j,\sigma_{b_j}^2 I_p)$.
Writing $\nu=(\nu_1^T,...,\nu_k^T)^T$ and $\rho=(\rho_1^T,...,\rho_n^T)^T$, the set of unknown parameters is $\theta=(\beta^T,\nu^T,\rho^T,{\sigma_a^2}^T,{\sigma_b^2}^T,{\sigma^2}^T,\delta^T,z^T)^T$. 
We replace $q(a)$ and $q(b)$ in the variational approximation with 
$q(\rho)=\prod_{i=1}^n q(\rho_i)$ and $q(\nu)=\prod_{j=1}^k q(\nu_j)$, where 
$q(\rho_i)$ is $N(\mu_{\rho_i}^q,\Sigma_{\rho_i}^q)$ for $i=1,...,n$, and
$q(\nu_j)$ is $N(\mu_{\nu_j}^q,\Sigma_{\nu_j}^q)$ for $j=1,...,k$. 
The variational algorithms with partial centering and full centering reparametrizations are known as `Algorithm 2' and `Algorithm 3' respectively. The variational lower bounds and parameter updates can be computed as before and can be found in the supplementary materials. The variational posterior for $\delta$ can be relaxed to be a normal distribution at convergence and similar adjustments (discussed in Section 3) apply to the variational lower bounds for Algorithms 2 and 3.

\section{Variational Greedy Algorithm}

In the greedy algorithm, VA refers to Variational Algorithm which can be 
Algorithm 1, 2 or 3. Let $f_k$ denote the $k$-component mixture model and $C_k$ the 
set of $k$ components that form the mixture model $f_k$. 
The greedy learning procedure can be outlined as follows.

\begin{enumerate}
\item Compute the one-component mixture model $f_1$ using VA.
\item Find the optimal way to split each of the components in the current mixture $f_k$.
This is done in the following manner. For each component $c_{j^*} \in C_k$, form  
$A_{j^*}=\{i \in \{1,...,n\}:j^*=\mbox{arg} \max_{1 \leq j \leq k} q_{ij}\},$
where $\{q_{ij}, 1 \leq i \leq n, 1 \leq j \leq k\}$ are the variational posterior probabilities of $f_k$. For $m=1,...,M$,
\begin{itemize}
\item randomly partition $A_{j^*}$ into two disjoint subsets $A_{{j1}^*}$ and $A_{{j2}^*}$ and form a $(k+1)$-component mixture by splitting the variational posterior probabilities of $c_{j^*}$ according to $A_{{j1}^*}$ and $A_{{j2}^*}$. That is, we create two subcomponents
$c_{{j1}^*}$ and $c_{{j2}^*}$ such that for $c_{{jl}^*}$, $q_{ij}$ is equal to 
the variational posterior probabilities of $c_{j^*}$ in $f_k$ if the $i$th observation 
lies in $A_{{jl}^*}$ and zero otherwise, $l=1,2$. The variational parameters of $c_{{j1}^*}$ and $c_{{j2}^*}$ required for initialization of the VA are set as equal to that of $c_{j^*}$. 
\item Variational parameters of all other components are set as those in $f_k$.
In the application of the VA, we do not update the variational parameters 
of components in $C_k-c_{j^*}$ as we are only interested in learning the optimal 
way of splitting $c_{j^*}$. Hence, we apply only a `partial' VA during this search. 
\end{itemize}  
      For each component $c_{j^*} \in C_k$, choose the run with the highest attained lower bound 
      among $M$ runs as that yielding the optimal way of splitting $c_{j^*}$. Let $L_{j^*}$ and 
      $f_{j^*}^{split}$ denote the lower bound and $(k+1)$-component mixture model respectively 
      corresponding to the optimal way of splitting $c_{j^*}$.
\item The components in $C_k$ are sorted in descending order according to $L_{j^*}$ and 
      then split in this order. After the $l$th split, the total number of components in 
      the mixture is ${k+l}$. Let $f_{k+l}^{temp}$ denote the mixture model obtained after
      $l$ splits. Suppose that at the $(l+1)th$ split, the component in $C_k$
      being split is $c_{j^*}$. We apply a `partial' VA again, keeping fixed variational parameters of components awaiting to be split. For the initialization, we set the variational parameters of $c_{{j1}^*}$ and $c_{{j2}^*}$ to be equal to those in 
      $f_{j^*}^{split}$ and the remaining variational parameters to be equal to those 
      in $f_{k+l}^{temp}$ if $l>1$ and $f_{j^*}^{split}$ if $l=1$. A split is considered `successful' if
      the estimated log marginal likelihood increases after the split. This process of splitting
      components is terminated once we encounter an unsuccessful split.
\item If the total number of successful splits in step 3 is $s$, then a $(k+s)$-component model $f_{k+s}^{temp}$ is obtained at the end of step 3. We apply VA on $f_{k+s}^{temp}$ until convergence updating all variational parameters this time to obtain mixture model $f_{k+s}$.
\item Repeat steps 2--3 until all splits of the current mixture model are unsuccessful.
\end{enumerate}

We have experimented with several dissimilarity measures based on Euclidean distance as well as variability-weighted similarity measures (Yeung {\it et al.}, 2003) in the case of repeated data to partition $A_{j^*}$ in step 2. Generally, VGA performed better when a random partition was used. Methods such as $k$-means clustering are also difficult to apply when there is missing data. The partitioning of $A_{j^*}$ into two disjoint subsets in step 2 serves only as an initialization to the `partial' VA to be carried out in search of the optimal way to split component $c_{j^*}$. If we consider an outright partitioning of the data by assigning observation $i$ to the $j^*$th component if $j^*=\mbox{arg} \max_{1 \leq j \leq k} q_{ij}$ where $\{q_{ij}, 1 \leq i \leq n, 1 \leq j \leq k\}$ are the variational posterior probabilities, it is still possible for observations originally from different components to be placed in the same component again in steps 3 and 4. This is due to the updating of the variational posterior probabilities $q_{ij}$ of all components which have been split in step 3 and that of all existing components in step 4.

The amount of computation is greatly reduced by the use of a `partial' VA 
as the algorithm converges quickly when the variational parameters of all other components
(except for the two subcomponents arising from the component being split) are fixed. As we are using only 
the run with the highest attained lower bound out of $M$ runs, it is not computationally efficient 
to continue every run to full convergence and we suggest using `short runs' in this search step. 
In later examples, we terminate each of these $M$ runs once the increment in the lower bound is 
less than 1. Suppose we are trying to split a component $c_{j^*}$ into two subcomponents
$c_{{j1}^*}$ and $c_{{j2}^*}$. After applying `partial' VA, the variational posterior probabilities of one of the two subcomponents sometimes reduce to zero for all of $i=1,...,n$, so that the two subcomponents effectively reduced to one. When this happens on the attempt leading to the highest variational lower bound among all $M$ attempts to split $c_{j^*}$, we suggest omitting $c_{j^*}$ in future splitting tests. This reduces the number of components we need to test for splitting and can be very useful when the number of components grows to a large number. For the examples discussed in this paper, we set $M$ to be 5 and the variational algorithm is deemed to have converged when the absolute relative change in the lower bound is less than $10^{-5}$. We note that the number of mixture components returned by the VGA may vary due to the random partitions in step 2 although the variation is relatively small compared to the number of clusters returned. The biggest advantage of the VGA is that it performs parameter estimation and model selection simultaneously and automatically returns a plausible number of components. It is possible however for the VGA to overestimate the number of components and some optional merge moves may be carried out after the VGA has converged if the user finds certain clusters to be very similar. This can be done quickly using a partial `VA' in which the variational parameters of all other components except the two to be merged are fixed. A merge move is considered `successful' if the estimated log marginal likelihood increases when two components are merged. While the VGA has been applied repeatedly in the examples for the purpose of analysing its performance, the user need only apply it once and may consider some merge moves if he finds clusters which are very similar. If multiple applications are used, we suggest using the estimated log marginal likelihood as a guideline to select the clustering solution. While reparametrizations using hierarchical centering increases the efficiency of the VGA, we have not observed that the number of components returned differs significantly due to the reparametrization. 

\section{Rate of convergence of variational approximation}
In this section, we show that the approximate rate of convergence of the variational algorithm by 
Gaussian approximation is equal to that of the corresponding Gibbs sampler. As reparametrizations using hierarchical centering can lead to improved convergence in the Gibbs sampler,
this result lends insight into how such reparametrizations can increase the efficiency of 
variational algorithms in the context of MLMMs. This is because the joint posterior
of the fixed and random effects in a LMM is Gaussian (with Gaussian priors
and Gaussian random effects distributions) when the variance parameters are known. 

Let the complete data be $Y_{aug}=(Y_{obs},Y_{mis})$ where $Y_{obs}$ is the observed data and $Y_{mis}$ is the missing data. 
Let the complete data likelihood be $p(Y_{aug}|\theta)$ where $\theta$ is a $p \times 1$ 
vector and $Y_{mis}$ $r \times 1$.
Suppose the prior for $\theta$ is $p(\theta)\propto 1$ and the target distribution is
$p(\theta,Y_{mis})= N\left( \left( \begin{smallmatrix} \mu_1 \\ \mu_2 \end{smallmatrix} \right),\Sigma \right)$ where $\Sigma=\left( \begin{smallmatrix} \Sigma_{11} & \Sigma_{12} \\ \Sigma_{21} & \Sigma_{22} \end{smallmatrix} \right)$. Let $ H=\Sigma^{-1}=\left( \begin{smallmatrix} H_{11} & H_{12} \\ H_{21} & H_{22} \end{smallmatrix} \right)$. It can be shown that 
\begin{align}
p(Y_{mis}|\theta,Y_{obs}) &= N\left(\mu_2-H_{22}^{-1} H_{21} (\theta-\mu_1),H_{22}^{-1}\right) \;\mbox{and} \nonumber\\ 
p(\theta|Y_{mis},Y_{obs}) &= N\left( \mu_1-H_{11}^{-1} H_{12} (Y_{mis}-\mu_2),H_{11}^{-1} \right). \nonumber
\end{align}
Sahu and Roberts (1999) showed that under such conditions, the rate of convergence of the EM algorithm
alternating between the two components $\theta$ and $Y_{mis}$ is equal to the rate of convergence  
of the corresponding two-block Gibbs sampler.
This rate is given by $\rho(B^{EM})$, where $ B^{EM}=H_{11}^{-1} H_{12} H_{22}^{-1} H_{21}$
and $\rho(.)$ denotes the spectral radius of a matrix. 

In the variational approach, we seek an approximation $q(\theta,Y_{mis})$ to the true posterior $p(\theta,Y_{mis}|Y_{obs})$ for which the KL divergence between $q$ and $p(\theta,Y_{mis}|Y_{obs})$ is minimized subject to the restriction that $q(\theta,Y_{mis})$ can be factorized as $q(\theta)q(Y_{mis})$.  The optimal densities are
\begin{align} 
q(Y_{mis}) & = N\left(\mu_2-H_{22}^{-1} H_{21} (\mu_{\theta}^q-\mu_1),\;H_{22}^{-1} \right) \;\mbox{and} \nonumber\\
q(\theta) & = N\left(\mu_1-H_{11}^{-1} H_{12} (\mu_{Y_{mis}}^q-\mu_2),\;H_{11}^{-1} \right), \nonumber
\end{align}
where $\mu_{\theta}^q$ and $\mu_{Y_{mis}}^q$ denote the mean of $q(\theta)$ and $q(Y_{mis})$ respectively. 
Starting with some initial estimate for $\mu_{\theta}^q$, we can iteratively update the 
parameters $\mu_{\theta}^q$ and $\mu_{Y_{mis}}^q$ until convergence. Let ${\mu_{\theta}^q}^{(t)}$ and $\mu_{Y_{mis}}^{q(t)}$ denote the $t$th iterates. It can be shown that  
\begin{gather*}
\mu_{Y_{mis}}^{q(t+1)} = H_{22}^{-1} H_{21} H_{11}^{-1} H_{12} \: \mu_{Y_{mis}}^{q(t)} 
+ \left(I_r-H_{22}^{-1} H_{21} H_{11}^{-1} H_{12}\right) \mu_2 \;\;\;\;\text{and} \\
{\mu_{\theta}^q}^{(t+1)} = B^{EM} {\mu_{\theta}^q}^{(t)} + \left(I_p-B^{EM}\right)\mu_1. 
\end{gather*}
The matrix rate of convergence of an iterative algorithm for which $\theta^{(t+1)}=M(\theta^{(t)})$ and $\theta^*$ is the limit is given by $DM(\theta^*)$ where $DM(\theta) = (\frac{\partial M_j(\theta)}{\partial \theta_i})$. A measure of the actual observed rate of convergence is given by the largest eigenvalue of $DM(\theta^*)$ (Meng, 1994). The rate of convergence of $\mu_{\theta}^q$ is therefore $\rho(B^{EM})$.
Since $H_{22}^{-1} H_{21} H_{11}^{-1} H_{12}$ and $B^{EM}$ share the same eigenvalues, 
the rate of convergence of $\mu_{Y_{mis}}^q$ is also $\rho(B^{EM})$.
The overall rate of convergence of the variational algorithm is thus $\rho(B^{EM}$).

Suppose we impose a tougher restriction on $q(\theta,Y_{mis})$. 
For a partition of $\theta$ into $m$ groups such that
$\theta=(\theta_1, ... , \theta_m)$ with $\theta_i$ a $r_i \times 1$ 
vector and $\sum{r_i}=p$, we assume that $q(\theta,Y_{mis})$
can be factorised as $\prod_{i=1}^m q(\theta_i) q(Y_{mis})$.
The optimal density of $q(Y_{mis})$ remains unchanged. 
Let $\mu_1 = (\mu_{11}, ... ,\mu_{1m}) $ and 
$$H_{11} = \left( \begin{smallmatrix} \Lambda_{11} & \Lambda_{12} & ... & \Lambda_{1m} \\
                                \Lambda_{21} & \Lambda_{22} & ... & \Lambda_{2m} \\
                                \vdots       & \vdots       & \ddots     & \vdots  \\
                                \Lambda_{m1} & \Lambda_{m2} & ... & \Lambda_{mm} \end{smallmatrix} \right).$$
be partitioned according to $\theta=(\theta_1, ... , \theta_m)$. The optimal density of $q(\theta_i)$ is
$$N( \mu_{1i}-\Lambda_{ii}^{-1}\sum_{j\neq i} {\Lambda_{ij}(\mu_{\theta_j}^q-\mu_{1j})} -\Lambda_{ii}^{-1} H_{12}(\mu_{Y_{mis}}^q-\mu_2),\Lambda_{ii}^{-1})\;\;\;\text{for}\;\;\;i=1,...,m.$$ 
This leads to the following iterative scheme. After initializing $\mu_{\theta_i}^q$, 
$i=1,...,m.$, we cycle though updates: 
\begin{itemize}
\item  $\mu_{Y_{mis}}^q  \leftarrow \mu_2-H_{22}^{-1} H_{21} (\mu_{\theta}^q-\mu_1) $
\item  $\mu_{\theta_i}^q \leftarrow \mu_{1i}-\Lambda_{ii}^{-1}\sum_{j\neq i} {\Lambda_{ij}(\mu_{\theta_j}^q-\mu_{1j})} -\Lambda_{ii}^{-1} H_{12}(\mu_{Y_{mis}}^q-\mu_2),\;i=1,...,m, $
\end{itemize} 
till convergence. Consider the $(t+1)th$ iteration. For notational simplicity, we replace 
$(\mu_{\theta_i}^{q(t)}-\mu_{1i})$ by $\lambda_{\theta_i}^{q(t)}$, $(\mu_{\theta}^{q(t)}-\mu_1)$ by $\lambda_{\theta}^{q(t)}$ and $(\mu_{Y_{mis}}^{q(t)}-\mu_2)$ by $\lambda_{Y_{mis}}^{q(t)}$.
Since $\lambda_{Y_{mis}}^{q(t+1)} = -H_{22}^{-1} H_{21}{\lambda_{\theta}^q}^{(t)}$, we have  
$$\left( \begin{matrix} 
\Lambda_{11} & 0            & ...      & 0      \\
\Lambda_{21} & \Lambda_{22} & ...      & 0      \\
\vdots       & \vdots       & \ddots   & \vdots \\
\Lambda_{m1} & \Lambda_{m2} & ...      & \Lambda_{mm} \end{matrix} \right)
\left( \begin{matrix} \lambda_{\theta_1}^{q(t+1)} \\ 
\lambda_{\theta_2}^{q(t+1)} \\ 
\vdots \\ 
\lambda_{\theta_m}^{q(t+1)}\end{matrix} \right)
+\left( \begin{matrix}   
0      & \Lambda_{12} & ... & \Lambda_{1m} \\
0      &       0      & ... & \Lambda_{2m} \\
\vdots & \vdots       & \ddots     & \vdots  \\
0      & 0 & ... & 0 \end{matrix} \right)
\left( \begin{matrix} \lambda_{\theta_1}^{q(t)} \\ 
\lambda_{\theta_2}^{q(t)} \\ 
\vdots \\ 
\lambda_{\theta_m}^{q(t)} \end{matrix} \right)
= H_{11}B^{EM}{\lambda_{\theta}^q}^{(t)}.$$
Let $L$ be the lower triangular matrix of $H_{11}$ and $U=L-H_{11}$. Then 
\begin{align*}
L{\lambda_{\theta}^q}^{(t+1)} -U{\lambda_{\theta}^q}^{(t)} &= H_{11}B^{EM}{\lambda_{\theta}^q}^{(t)} \\
\Leftrightarrow {\lambda_{\theta}^q}^{(t+1)} &= L^{-1}U{\lambda_{\theta}^q}^{(t)}+L^{-1}(L-U)B^{EM}{\lambda_{\theta}^q}^{(t)}  \\
\Leftrightarrow {\lambda_{\theta}^q}^{(t+1)} &= [B_{aug}+(I_p-B_{aug})B^{EM}]{\lambda_{\theta}^q}^{(t)} 
\end{align*} 
where $B_{aug}=L^{-1}U$. Therefore the rate of convergence of $\lambda_{\theta}^q$ and hence, that of $\mu_{\theta}^q$ 
is $\rho(B_{aug}+(I_p-B_{aug})B^{EM})$. As the rate of convergence $r$, is defined as 
$r=\lim_{t\rightarrow \infty} \frac{\left\|\theta^{(t+1)}-\theta^*\right\|}{\left\|\theta^{(t)}-\theta^*\right\|},$
the rate of convergence of $\lambda_{Y_{mis}}^q$ and hence $\mu_{Y_{mis}}^q$ is given by 
\begin{gather*}
\lim_{t\rightarrow \infty} \frac{\left\| {\lambda_{Y_{mis}}^{q(t+1)}}-{\lambda_{Y_{mis}}^{q*}}\right\|}
{\left\| {\lambda_{Y_{mis}}^{q(t)}}-{\lambda_{Y_{mis}}^{q*}}\right\|} = 
\lim_{t\rightarrow \infty} \frac{\left\| -H_{22}^{-1} H_{21}{\lambda_{\theta}^q}^{(t)}+H_{22}^{-1} H_{21}{\lambda_{\theta}^q}^*\right\|}
{\left\| -H_{22}^{-1} H_{21}{\lambda_{\theta}^q}^{(t-1)}+H_{22}^{-1} H_{21}{\lambda_{\theta}^q}^*\right\|} = \lim_{t\rightarrow \infty} \frac{\left\| {\lambda_{\theta}^q}^{(t)}-{\lambda_{\theta}^q}^*\right\|}
{\left\| {\lambda_{\theta}^q}^{(t-1)}-{\lambda_{\theta}^q}^*\right\|}  \nonumber 
\end{gather*}
which is equal to the rate of convergence of $\mu_{\theta}^q$. The overall rate of convergence of the variational algorithm is thus $\rho(B_{aug}+(I_p-B_{aug})B^{EM})$ which is equal to the rate of convergence of the Gibbs sampler that sequentially updates components of $\theta$, and then block updates $Y_{mis}$ derived by Sahu and Roberts (1999). Although the theory developed may not be directly applicable to LMMs with unknown variance components as well as MLMMs in general, it suggests to consider hierarchical centering in the context of variational algorithms and our examples show that there is some gain in efficiency due to the reparametrizations.

\section{Examples}
To illustrate the methods proposed, we apply VGA using Algorithms 1, 2 and 3 on three real data sets (application of Algorithm 2 on yeast galactose data set can be found in supplementary materials). We also consider simulated data sets in Section 7.3 where VGA is compared with EMMIX-WIRE (Ng {\it et al.}, 2006). In Section 7.2, we report the gain in efficiency from reparametrization of the model using hierarchical centering. In the examples below, an outright partitioning of the data is obtained by assigning observation $i$ to the $j^*$th component if $j^*=\mbox{arg} \max_{1 \leq j \leq k} q_{ij}$, where $\{q_{ij}, 1 \leq i \leq n, 1 \leq j \leq k\}$ are the variational posterior probabilities of the mixture model obtained using VGA.

\subsection{Clustering of time course data}
Using DNA microarrays and samples from yeast cultures synchronized by three independent methods, Spellman {\it et al.} (1998) identified 800 genes that meet an objective minimum criterion for cell cycle regulation.
 We consider the 18 $\alpha$-factor synchronization where the yeast cells were sampled at 7 min intervals for 119 mins and a subset of 612 genes that have no missing gene expression data across all 18 time points. This data set was previously analyzed by Luan and Li (2003) and Ng {\it et al.} (2006) and is available online from the yeast cell cycle analysis project at http://genome-www.stanford.edu/cellcycle/. Our aim is to obtain an optimal clustering of these genes using the VGA. Following Ng {\it et al.} (2006), we take $n=612$ genes, $W_i=1_{18}$, $V_i=I_{18}$, $u_i=1$ and $X_i$ to be an $18 \times 2$ matrix with the $(l+1)$th row ($l=0,...,17$) as $(\cos(2\pi(7l)/\omega), \sin(2\pi(7l)/\omega),$ where $\omega=53$ is the period of the cell cycle for $i=1,...,n$. For the error terms, we take $g=1$ and $\kappa_{i1}=18$ for $i=1,...,n$ so that the error variance of each mixture component is constant across the 18 time points. We used the following priors, $\delta\sim N(0,1000I)$, $\beta_j\sim N(0,1000I)$ for $j=1,...,k,$ and $IG(0.01,0.01)$ for $\sigma_{a_j}^2$, $\sigma_{b_j}^2$, $j=1,...,k$ and $\sigma_{jl}^2$,
$j=1,...,k$, $l=1,...,g$. 

\begin{figure}
\begin{center}
\includegraphics[width=160mm]{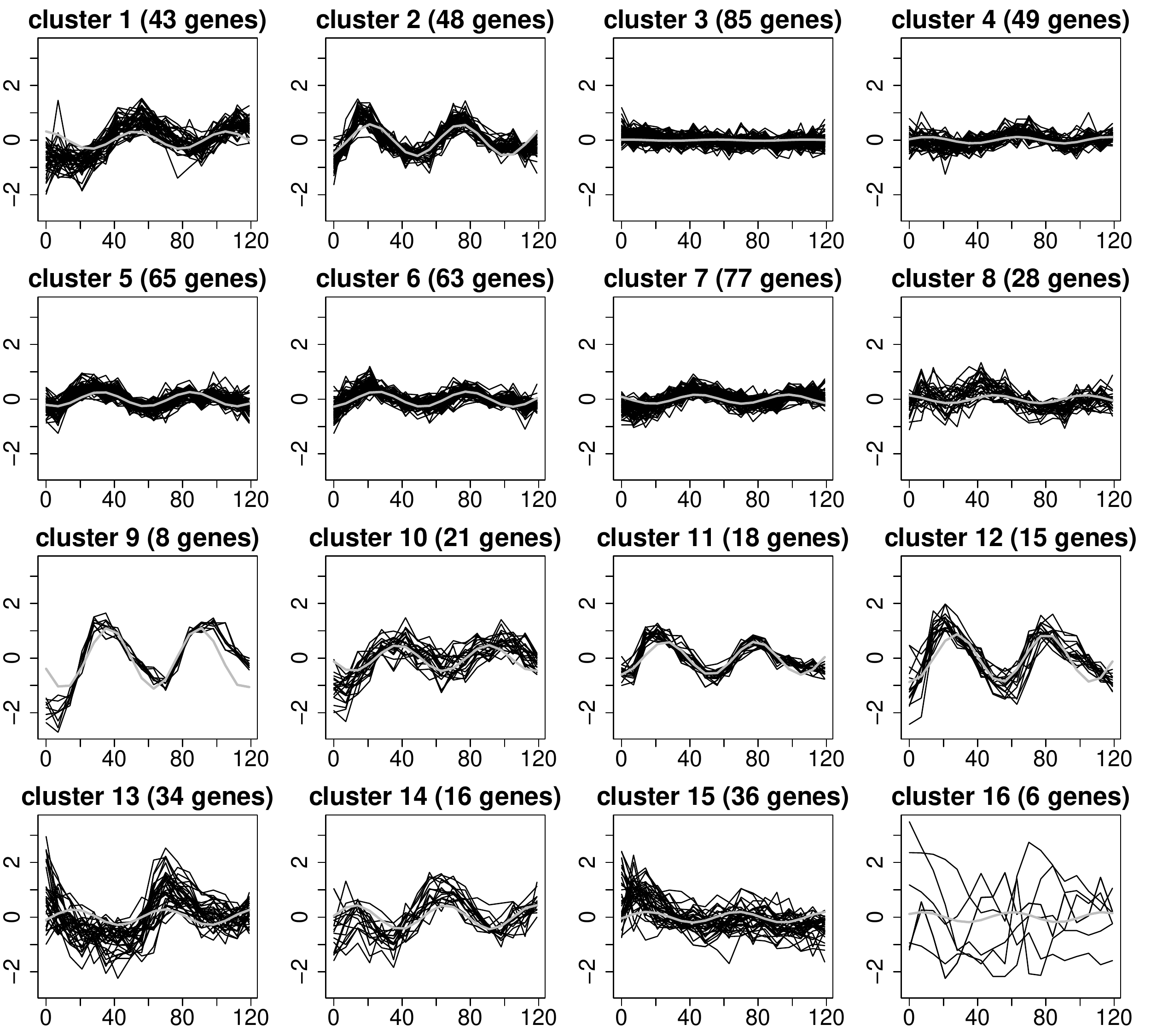} 
\end{center}
\caption{\label{time} Clustering results for time course data obtained from applying the VGA using Algorithm 1. The $x$-axis are the time points and $y$-axis are the gene expression levels. Line in grey is the posterior mean of the fixed effects given by $X_i \mu_{\beta_j}^q$. }
\end{figure}

Applying the VGA using Algorithm 1 ten times, we obtained a 15-component mixture once, a 16-component mixture six times and a 17-component mixture thrice. The mode is 16 and we report the clustering for a 16-component mixture obtained from the VGA in Figure \ref{time}. 
For this clustering, we attempted several merge moves on clusters which appear similar such as 3 with 4, 5 with 6, 7 with 8 and 13 with 14. 
These merge moves did not result in a higher estimated log marginal likelihood. However, we observed that cluster 2 (48 genes) was split into two clusters in one of the 17-component mixture models and these two clusters can be merged successfully with a higher estimated log marginal likelihood being obtained. 
Thus, it is possible for the VGA to overestimate the number of mixture components and merge moves can be considered when similar clusters are encountered. We note however that the variation in the number of mixture components returned by the VGA is relatively small. For this data set, the number of clusters returned by VGA was generally larger than that obtained by Ng {\it et al.} (2006) where BIC was used for model selection and the optimal number of clusters was reported as 12. Any interpretation of the differences in results would need to be pursued with the help of subject matter 
experts, but our later simulation studies tend to indicate that BIC underestimates
the true model so that possibly our clustering is preferable from this point of view.
Of course it may be argued that the ability to estimate the `true model' is not a 
chief concern in clustering applications where interpretability of the results in 
the substantive scientific context is the primary motivation. 

\subsection{Clustering of water temperature data}
\begin{figure}
\begin{center}
\includegraphics[width=160mm]{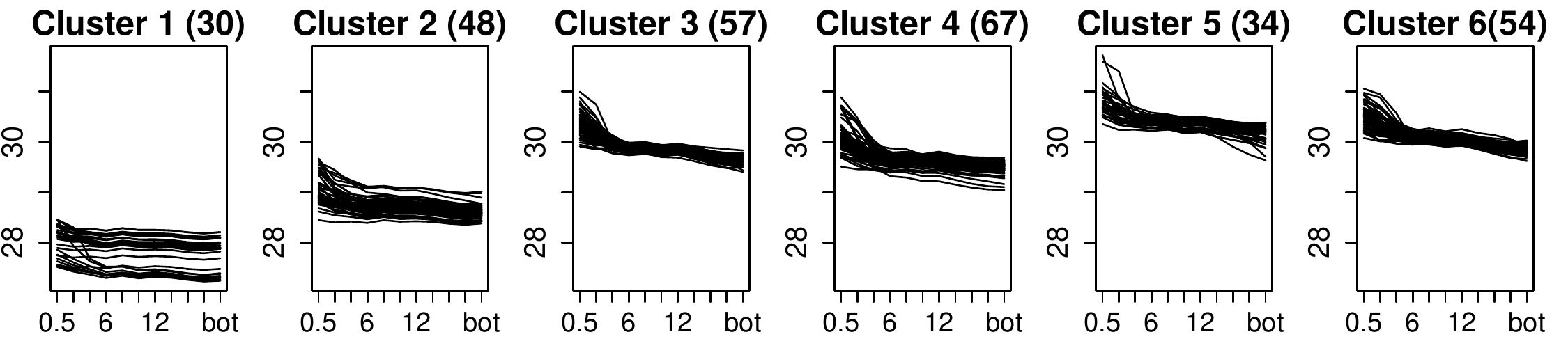} 
\end{center}
\caption{\label{water} Clustering results for water temperature data. The $x$-axis is the depth and $y$-axis is the water temperature.}
\end{figure}
\begin{figure}
\begin{center}
\includegraphics[width=160mm]{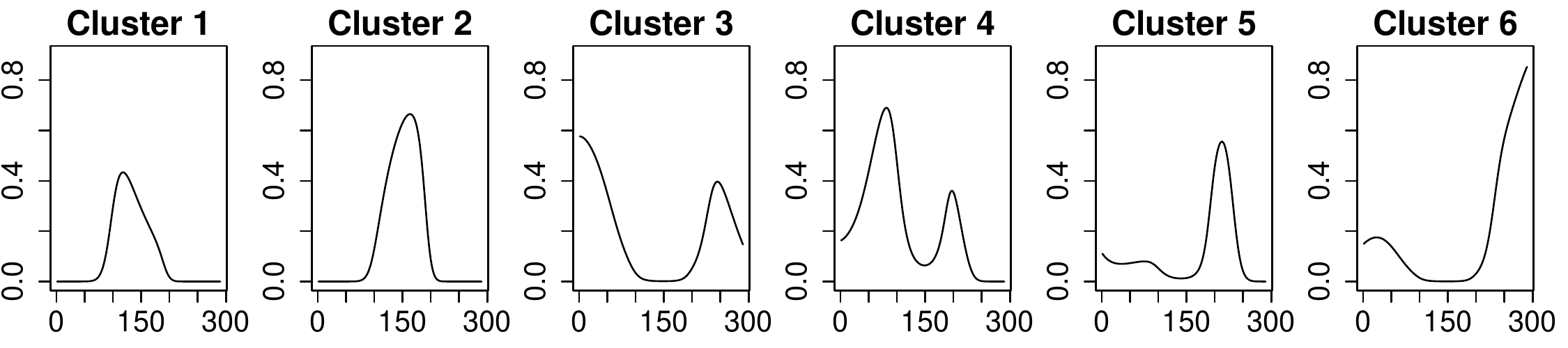} 
\end{center}
\caption{\label{gatingprob} Fitted probabilities by gating function for clusters 1 to 6. The $x$-axis are days numbered 1 to 290 and y-axis are the probabilities.}
\end{figure}

We consider the daily average water temperature readings during the period 
9 September 2010--10 August 2011 collected at a monitoring station at Upper Peirce 
Reservoir, Singapore. No data were available during the periods 
23 December 2010--28 December 2010, 10 February 2010--23 February 2010 and 
14 April 2011--10 May 2011. Readings were collected at eleven depths from the water surface;
0.5m, 2m, 4m, 6m, 8m, 10m, 12m, 14m, 16m, 18m and at the bottom.
Using data from the remaining 290 days, we apply the VGA to obtain a clustering of this data. 
We take $n=290$, $n_i=11$ and $X_i=W_i=V_i=I_{11}$ for $i=1,...,n$. 
We set $g=11$ with $\kappa_{il}=1$ for $i=1,...,n$, $l=1,...,g$ so that the 
error variance of each mixture component is allowed to be different
at different depths. For the mixture weights, we set
$u_i=(1,i,i^2,i^3)$, $i=1,...,n,$ and subsequently standardize columns 2--4 in the matrix
$U=(u_1^T,...,u_n^T)^T$ to take values between -1 and 1, centered at 0. 
We used the following priors, $\delta\sim N(0,1000I)$,
$\beta_j\sim N(0,10000I)$ for $j=1,...,k,$ and $IG(0.01,0.01)$ 
for $\sigma_{a_j}^2$, $\sigma_{b_j}^2$, $j=1,...,k$ and $\sigma_{jl}^2$,
$j=1,...,k$, $l=1,...,g$.
Applying VGA with Algorithm 3 five times, we obtained a 6-component model each time with very similar results. The clustering of a 6-component fitted model is shown in Figure \ref{water} and the fitted 
probabilities from the gating function are shown in Figure \ref{gatingprob}.
For comparison, we apply VGA with Algorithm 1 five times. A 6-component mixture
model was obtained on all five attempts. The average CPU time taken to fit a 6-component model 
using VGA with Algorithm 1 was 2114 seconds compared to 932 seconds by Algorithm 3. In this example,
hierarchical centering reparametrization has helped to improve the rate of convergence with the computation time reduced by more than half. The Upper Peirce Reservoir uses aeration devices intended to mix 
the water at different depths, with the aim of controlling outbreaks 
of phytoplankton and algal scums. On days when these aeration devices 
are operational, it is expected that there will be less stratification 
of the temperature with depth. Accurate records of the operation of 
the aeration devices were not available to us and there is some interest 
in seeing whether the clusters divide into more or less stratified 
components giving some insight into when the aeration devices were used.

\subsection{Simulation study}
We report results from a simulation study in which VGA is compared with 
EMMIX-WIRE developed by Ng {\it et al.} (2006). EMMIX-WIRE fits MLMMs by likelihood 
maximization using the EM algorithm and is able to handle the clustering of correlated data 
that may be replicated. We compare the performance of EMMIX-WIRE with VGA using 10 data sets simulated from model (\ref{MLMM}). Each data set consist of $n=499$ vectors of 
dimension $n_i=18$ and each contain 12 clusters of sizes 43, 48, 85, 49, 65, 77, 8, 21, 18, 15, 34 and 36. These clusters are based on the 16-component mixture model in Figure 
\ref{time} fitted to the time course data in Section 7.1 from which 12 distinctive clusters have been selected. In particular, we have left out clusters 6, 8, 14 and 16. The values of the unknown parameters $\beta_j$, $\sigma_{a_j}^2$, $\sigma_{b_j}^2$, $j=1,...,k,$ and $\sigma_{jl}^2$, $j=1,...,k$, 
$l=1,...,g$ in model (\ref{MLMM}) were taken to be equal to the variational posterior mean values of 
the 16-component mixture in Figure \ref{time} and $g=1$. The design matrices $X_i$, $W_i$ and $V_i$, 
$i=1,...,n,$ are as described in Section 7.1 and we used the same priors as before.

For each of the 10 data sets, we ran EMMIX-WIRE with the number of components 
ranging from 6 to 15 and used the BIC for model selection.
The optimal number of components is taken to be that which minimizes 
$-2\log(Lik)+(par)\log n,$ where $par$ denotes the number of parameters in the model and 
$Lik$ is the likelihood. We used the approximation of $Lik$ from the output of 
EMMIX-WIRE for the computation of the BIC. See Ng {\it et al.} (2006) for details on how 
the likelihood was approximated. We ran EMMIX-WIRE again, this time fixing the number of 
components as 12. We also applied the VGA with Algorithm 1 once for each of the 10 
data sets. The adjusted Rand Index (ARI) (Hubert and Arabie, 1985) for the clustering of the fitted model relative to the true grouping of all 499 observations into 12 clusters was then computed in each case.
The results are summarized in Table \ref{simu}.
\begin{table}
\begin{center}
\begin{small}
\begin{tabular}{cccccc}
        & \multicolumn{3}{c}{EMMIX-WIRE}               &\multicolumn{2}{c}{VGA}\\ \hline
        &                  & Optimal    & 17-comp      &                 & Optimal\\
Data    & No. of clusters  & model      & model        & No. of clusters & model \\
set     & in optimal model & ARI        & ARI          & in optimal model& ARI   \\ \hline
1       & 8                & 0.658      & 0.725        & 12              & 0.966 \\ 
2       & 8                & 0.606      & 0.837        & 12              & 0.898 \\ 
3       & 8				   & 0.534		& 0.724		   & 11			     & 0.774 \\
4       & 9                & 0.774		& 0.808		   & 13              & 0.928 \\
5       & 8                & 0.604      & 0.724        & 12              & 0.951 \\
6       & 7                & 0.545      & 0.904        & 12              & 0.951 \\
7       & 7                & 0.500      & 0.697        & 11              & 0.779 \\
8       & 8                & 0.649      & 0.642        & 12              & 0.888 \\ 
9       & 6				   & 0.522      & 0.537        & 11              & 0.755 \\
10      & 6                & 0.485      & 0.684        & 12              & 0.922 \\ \hline
\end{tabular}
\end{small}
\end{center}
\caption{\label{simu} Simulation results comparing EMMIX-WIRE with VGA.}
\end{table}
From Table \ref{simu}, the ARI attained by VGA was consistently higher than 
that attained by EMMIX-WIRE. It is also interesting to note that in almost 
all the ten sets of simulated data, the ARI attained by the 12-component model 
fitted by EMMIX-WIRE was higher than that attained by the optimal model 
identified by BIC. So BIC tends to underestimate the number of components
here, although the implications of this for applications in clustering algorithms 
may be less clear.

\section{Conclusion}
We have proposed fitting MLMMs with variational methods and developed an 
efficient VGA which is able to perform parameter estimation and model 
selection simultaneously. This greedy approach handles initialization 
automatically and returns a plausible value for the number of mixture 
components. The experiments we have conducted showed that the VGA does 
not systematically underestimate nor overestimate the number of mixture 
components. For the simulated data sets considered, VGA was able to
return mixture models where the number of mixture components is very
close to the correct number of components. We further showed empirically
that hierarchical centering can help to improve the rate of convergence 
in variational algorithms significantly. Some theoretical support was 
also provided for this observation. Implementation of the VGA is straightforward as no further derivation is 
required once the basic variational algorithms are available. This greedy 
approach is not limited to MLMMs and could potentially be extended to 
fitting other models using variational methods. All code was written in
the R language and run on a dual processor Window PC 3GHz workstation.

\section{Supplementary materials}
The derivation of the variational lower bound in (\ref{lb_alg1}) and the expressions of the variational lower bounds and parameter updates for Algorithms 2 and 3 can be found in the supplementary materials. An example on application of Algorithm 2 to yeast galactose data is also included. 

\section{Acknowledgements}
Siew Li Tan was partially supported as part of the Singapore-Delft Water Alliance (SDWA)'s
tropical reservoir research programme. We thank SDWA for supplying the water temperature data set
and Dr David Burger and Dr Hans Los for their valuable comments and suggestions.

\begin{small}
\section*{References}
\bib
Armagan, A. and Dunson, D. (2011).
Sparse variational analysis of linear mixed models for large data sets.
{\it Statistics and Probability Letters}, 81, 1056--1062.

\bib
Attias, H. (1999).
Inferring parameters and structure of latent variable models by variational Bayes.
In {\it Proceedings of the 15th Conference on Uncertainty in Artificial Intelligence}, 21--30.

\bib
Blei, D.M. and Jordan, M.I. (2006).
Variational inference for Dirichlet process mixtures. 
{\it Bayesian Analysis}, 1, 121--144.

\bib
Braun, M. and McAuliffe, J. (2010).
Variational inference for large-scale models of discrete choice.
{\it Journal of the American Statistical Association}, 105, 324--335.

\bib
Biernacki, C.,  Celeux, G. and Govaert, G. (2003).
Choosing starting values for the EM algorithm for getting the highest likelihood in multivariate Gaussian mixture models.
{\it Computational Statistics and Data Analysis}, 41, 561--575.

\bib
Bishop, C.M. and Svens\'{e}n, M. (2003).
Bayesian hierarchical mixtures of experts.
In {\it Proceedings of the 19th Conference on Uncertainty in Artificial Intelligence}, 57--64. 

\bib
Booth, J.G., Casella, G. and Hobert, J.P. (2008).
Clustering using objective functions and stochastic search.
{\it Journal of the Royal Statistical Society: Series B}, 70, 119--139.

\bib
Celeux, G., Martin O. and Lavergne C. (2005).
Mixture of linear mixed models for clustering gene expression profiles from repeated microarray experiments.
{\it Statistical Modelling}, 5, 243--267.

\bib
Chen, M.H., Shao, Q.M. and Ibrahim, J.G. (2000).
{\it Monte Carlo methods in Bayesian computation}. Springer. 

\bib
Coke, G. and Tsao, M. (2010).
Random effects mixture models for clustering electrical load series. 
{\it Journal of Time Series Analysis}, 31, 451--464. 

\bib
Constantinopoulos, C. and Likas, A. (2007).
Unsupervised learning of Gaussian mixtures based on variational component splitting.
{\it IEEE Transactions on Neural Networks}, 18, 745--755.

\bib
Corduneanu, A, and Bishop, C.M. (2001).
Variational Bayesian model selection for mixture distributions.
In {\it Proceedings of 8th International Conference on Artificial Intelligence and Statistics}, 27--34.

\bib
Dempster, A.P., Laird, N.M. and Rubin, D.B. (1977).
Maximum likelihood from incomplete data via the EM algorithm. 
{\it Journal of the Royal Statistical Society: Series B}, 39, 1--38. 

\bib
Gelfand, A.E., Sahu, S.K. and Carlin, B.P. (1995). 
Efficient parametrisations for normal linear mixed models.
{\it Biometrika}, 82, 479--488.

\bib
Hubert, L. and Arabie, P. (1985).
Comparing partitions.
{\it Journal of Classification}, 2, 193--218.

\bib
Jacobs, R.A., Jordan, M.I., Nowlan, S.J. and Hinton, G.E. (1991). 
Adaptive mixtures of local experts. 
{\it Neural Computation}, 3, 79--87.

\bib
Jordan, M.I., Ghahramani, Z., Jaakkola, T.S., Saul, L.K. (1999).
An introduction to variational methods for graphical models. 
{\it Machine Learning}, 37, 183--233.

\bib
Luan, Y. and Li, H. (2003).
Clustering of time-course gene expression data using a mixed-effects model with B-splines.
{\it Bioinformatics}, 19, 474--482.

\bib
McGrory, C.A. and Titterington, D.M. (2007).  
Variational approximations in Bayesian model selection for finite mixture distributions.
{\it Computational Statistics and Data Analysis}, 51, 5352--5367.

\bib
McLachlan, G.J., Do, K.A. and Ambroise, C. (2004).
{\it Analyzing microarray gene expression data}. New York: Wiley.

\bib
Meng, X.L. (1994).
On the rate of convergence of the ECM algorithm.
{\it Annals of Statistics}, 22, 326--339. 

\bib
Ng, S.K., McLachlan, G.J., Wang, K., Ben-Tovim Jones, L. and Ng, S.-W. (2006).  
A mixture model with random-effects components for clustering correlated gene-expression profiles.
{\it Bioinformatics}, 22, 1745--1752. 

\bib
Nott, D.J., Tan, S.L., Villani, M. and Kohn, R. (2011). 
Regression density estimation with variational methods and stochastic approximation. 
{\it Journal of Computational and Graphical Statistics}, to appear. Preprint: http://www.mattiasvillani.com/wp-content/uploads/2011/07/varia \linebreak tional-heteroscedastic-moe-july-6-20114.pdf

\bib
Ormerod, J.T. and Wand, M.P. (2010).
Explaining variational approximations.
{\it The American Statistician}, 64, 140--153. 

\bib
Ormerod, J.T. and Wand, M.P. (2012).
Gaussian variational approximate inference for generalized linear mixed models.
{\it Journal of Computational and Graphical Statistics}, 21, 2--17.

\bib
Papaspiliopoulos, O., Roberts, G.O. and Sk\"old, M. 
A general framework for the parametrization of hierarchical models.
{\it Statistical Science}, 22, 59--73. 

\bib
Sahu, S.K. and Roberts, G.O. (1999).
On convergence of the EM algorithm and the Gibbs sampler.
{\it Statistics and Computing}, 9, 55--64.

\bib
Scharl, T., Gr\"{u}n, B. and Leisch, F. (2010).
Mixtures of regression models for time course gene expression data: evaluation of initialization and random effects.
{\it Bioinformatics}, 26, 370--377.

\bib
Schwarz, G. (1978). 
Estimating the dimension of a model.
{\it Annals of Statistics}, 6, 461--464.

\bib
Spellman, P.T., Sherlock, G., Zhang, M.Q., Iyer, V.R., Anders, K., Eisen, M.B., Brown, P.O., Botstein, D. and Futcher, B. (1998).
Comprehensive identification of cell cycle-regulated genes of the yeast Saccharomyces cerevisiae by microarray hybridization.
{\it Molecular Biology of the Cell}, 9, 3273--3297.

\bib
Ueda, N. and Ghahramani, Z. (2002).
Bayesian model search for mixture models based on optimizing variational bounds.
{\it Neural Networks}, 15, 1223--1241.

\bib
Verbeek, J.J., Vlassis, N. and Kr\"{o}se, B. (2003).
Efficient greedy learning of Gaussian mixture models.
{\it Neural Computation}, 15, 469--485.

\bib
Wand, M.P. (2002).  
Vector differential calculus in statistics. 
{\it The American Statistician}, 56, 55--62. 

\bib
Wang, B. and Titterington, D.M. (2005).  
Inadequacy of interval estimates corresponding to variational Bayesian approximations.
In {\it Proceedings of the 10th International Workshop on Artificial Intelligence}, 373--380.

\bib
Waterhouse, S., MacKay, D. and Robinson, T. (1996). 
Bayesian methods for mixtures of experts.  
{\it Advances in Neural Information Processing Systems 8}, 351--357.

\bib
Winn, J. and Bishop, C.M. (2005). 
Variational message passing. 
{\it Journal of Machine Learning Research}, 6, 661--694.

\bib
Wu, B., McGrory, C.A. and Pettitt, A.N. (2012).  
A new variational Bayesian algorithm with application to human mobility pattern modeling.
{\it Statistics and Computing}, 22, 185--203. 

\bib
Yeung, K.Y., Medvedovic, M. and Bumgarner, R.E. (2003).
Clustering gene-expression data with repeated measurements.
{\it Genome Biology}, 4, Article R34.
\end{small}
\end{document}


\renewcommand{\thefootnote}{\fnsymbol{footnote}}
\begin{center}
{\LARGE Supplementary materials for ``Variational approximation for mixtures of linear mixed models''\\}
\end{center}

\begin{center}
{\Large Siew Li Tan and David J. Nott} \\
\end{center}

\section*{Derivation of variational lower bound}

The variational lower bound can be written as $E_q(\log p(y,\theta))-E_q(\log q(\theta))$, where $E_q(\cdot)$ denotes the expectation with respect to $q$. Consider the first term, $E_q(\log p(y,\theta))$. Let $\zeta_{ij}=I(z_i=j)$ where $I(\cdot)$ denotes the indicator function. We have
\begin{align*}
  \log p(y,\theta) & = \sum_{i=1}^n \sum_{j=1}^k \zeta_{ij}\left\{ 
                       \log p(y_i|z_i=j,\beta_j,a_i,b_j,\Sigma_{ij})
  										 +\log p(a_i|\sigma_{a_j}^2)+\log p_{ij} \right\} +\log p(\delta) \\
               &\;\;\; +\sum_{j=1}^k \left\{ \log p(\beta_j)+ \log p(b_j|\sigma_{b_j}^2)
                       + \log p(\sigma_{a_j}^2) + \log p(\sigma_{b_j}^2) 
                       + \sum_{l=1}^g \log p(\sigma_{jl}^2) \right\}.
\end{align*}
Taking expectations with respect to $q$, we obtain 
\begin{align}
E_q(\log p(y,\theta)) &= \sum_{i=1}^n \sum_{j=1}^k q_{ij} \left\{ 
   - \sum_{l=1}^g \frac{\kappa_{il}}{2}(\log (\lambda_{jl}^q)-\psi(\alpha_{jl}^q))
   -\frac{1}{2} (\xi_{ij}^T {\Sigma_{ij}^q}^{-1} \xi_{ij}
   + \mbox{tr} ({\Sigma_{ij}^q}^{-1} \Lambda_{ij}))\right.  \nonumber \\
 &\;\;\; \left. -\frac{s_1}{2}(\log \lambda_{a_j}^q-\psi(\alpha_{a_j}^q))
   - \frac{\alpha_{a_j}^q}{2\lambda_{a_j}^q} \left( \mbox{tr} (\Sigma_{a_i}^q) 
   + {\mu_{a_i}^q}^T \mu_{a_i}^q  \right)+ \log p_{ij} \right\} +\log p(\mu_\delta^q) \nonumber \\
 &\;\;\; -\frac{1}{2}\sum_{j=1}^k \left\{\log |\Sigma_{\beta_j}|
   + \mbox{tr} (\Sigma_{\beta_j}^{-1} \Sigma_{\beta_j}^q) 
   + {\mu_{\beta_j}^q}^T \Sigma_{\beta_j}^{-1} \mu_{\beta_j}^q  
   +s_2(\log\lambda_{b_j}^q-\psi(\alpha_{b_j}^q)) \right. \nonumber \\
 &\;\;\; \left. +\frac{\alpha_{b_j}^q}{\lambda_{b_j}^q} (\mbox{tr} (\Sigma_{b_j}^q) 
   + {\mu_{b_j}^q}^T \mu_{b_j}^q)\right\} 
   + \sum_{j=1}^k \left\{ \alpha_{a_j} \log\lambda_{a_j} - \log \Gamma(\alpha_{a_j})
   - \frac{\lambda_{a_j} \alpha_{a_j}^q}{\lambda_{a_j}^q}  \right.  \nonumber \\
 &\;\;\; \left. -(\alpha_{a_j}+1)(\log \lambda_{a_j}^q-\psi(\alpha_{a_j}^q)) 
   + \alpha_{b_j} \log \lambda_{b_j}- \log \Gamma(\alpha_{b_j})
   - \frac{\lambda_{b_j} \alpha_{b_j}^q}{\lambda_{b_j}^q} \right. \nonumber \\
 &\;\;\; \left. -(\alpha_{b_j}+1)(\log \lambda_{b_j}^q-\psi(\alpha_{b_j}^q))
    + \sum_{l=1}^g \left[ \alpha_{jl} \log \lambda_{jl} 
    - \log \Gamma(\alpha_{jl})- \frac{\lambda_{jl} \alpha_{jl}^q}{\lambda_{jl}^q} \right. \right.\nonumber \\
 &\;\;\; \left. \left. -(\alpha_{jl}+1)\left(\log \lambda_{jl}^q-\psi(\alpha_{jl}^q)\right) \right] \right\}
    -\frac{N+k(p+s_2)+ns_1}{2}\log(2\pi),  \label{term1}
\end{align}
where $\Gamma(\cdot)$ and $\psi(\cdot)$ denote the gamma and digamma functions respectively, $p_{ij}$ is evaluated by setting $\delta=\mu_{\delta}^q$, $p(\mu_\delta^q)$ denotes the prior distribution for $\delta$ evaluated at $\mu_\delta^q$, $\xi_{ij}=y_i-X_i\mu_{\beta_j}^q-W_i\mu_{a_i}^q-V_i \mu_{b_j}^q$, ${\Sigma_{ij}^q}^{-1} = \text{blockdiag} \left(\frac{\alpha_{j1}^q}{\lambda_{j1}^q}I_{\kappa_{i1}}, ...,
\frac{\alpha_{jg}^q}{\lambda_{jg}^q}I_{\kappa_{ig}} \right)$ and
$\Lambda_{ij}=X_i\Sigma_{\beta_j}^qX_i^T+W_i\Sigma_{a_i}^q W_i^T+V_i \Sigma_{b_j}^q V_i^T$. 
Turning to the second term, $E_q(\log q(\theta))$, we have 
\begin{align}
E_q(\log q(\theta)) &=  \sum_{j=1}^k E_q\left\{\log q(\beta_j)+ \log q(b_j) 
                        + \log q(\sigma_{b_j}^2) + \log q(\sigma_{a_j}^2)
                        + \sum_{l=1}^g \log q(\sigma_{jl}^2)  \right\}    \nonumber \\
                    &\;\;\; +\sum_{i=1}^n E_q(\log q(a_i))
                            +\sum_{i=1}^n \sum_{j=1}^k q_{ij}\log q_{ij} \nonumber \\
& =  \sum_{j=1}^k \left\{ - \frac{1}{2}(\log |\Sigma_{\beta_j}^q|+\log |\Sigma_{b_j}^q|)
     + \alpha_{b_j}^q \log \lambda_{b_j}^q 
     -(\alpha_{b_j}^q+1)(\log\lambda_{b_j}^q-\psi(\alpha_{b_j}^q)) \right.\nonumber \\
&\;\;\; \left. - \log \Gamma(\alpha_{b_j}^q)- \alpha_{b_j}^q 
        +\alpha_{a_j}^q \log \lambda_{a_j}^q 
        -(\alpha_{a_j}^q+1)(\log \lambda_{a_j}^q-\psi(\alpha_{a_j}^q))
        - \log \Gamma(\alpha_{a_j}^q)  \right.    \nonumber \\
&\;\;\; \left. - \alpha_{a_j}^q + \sum_{l=1}^g \left[ \alpha_{jl}^q \log \lambda_{jl}^q 
        - \log \Gamma(\alpha_{jl}^q)- \alpha_{jl}^q 
        -(\alpha_{jl}^q+1)\left(\log \lambda_{jl}^q-\psi(\alpha_{jl}^q)\right) \right] \right\} \nonumber \\
&\;\;\; - \frac{1}{2} \sum_{i=1}^n \log |\Sigma_{a_i}^q|
        - \frac{k(p+s_2)+ns_1}{2}(\log(2\pi)+1) +\sum_{i=1}^n \sum_{j=1}^k q_{ij}\log q_{ij} \label{term2}
\end{align}
and putting (\ref{term1}) and (\ref{term2}) together gives the lower bound.

\section*{Algorithm 2 (partial centering when $X_i=W_i$)}
(Updates in steps 5, 7, 9, 10, 11 and 13 remain the same as in Algorithm 1 with $s_1=p$.)
Initialize: $q_{ij}$ for $i=1,...,n$, $j=1,...,k$,
            $\mu_{b_j}^q$, $\mu_{\beta_j}^q$, $\frac{\alpha_{a_j}^q}{\lambda_{a_j}^q}$ and
            $\frac{\alpha_{b_j}^q}{\lambda_{b_j}^q}$ for $j=1,...,k$,
            $\frac{\alpha_{jl}^q}{\lambda_{jl}^q}$ for $j=1,...,k$, $l=1,...,g$.
Do until the change in the lower bound between iterations is less than a tolerance:
\begin{enumerate}
\item $\Sigma_{\eta_i}^q \leftarrow \left( \sum_{j=1}^k q_{ij} \frac{\alpha_{a_j}^q}{\lambda_{a_j}^q} I_p 
		  + X_i^T (\sum_{j=1}^k q_{ij} {\Sigma_{ij}^q}^{-1}) X_i \right)^{-1}$.
\item $\mu_{\eta_i}^q \leftarrow \Sigma_{\eta_i}^q \sum_{j=1}^k q_{ij} 
      \left\{ \frac{\alpha_{a_j}^q}{\lambda_{a_j}^q} \mu_{\beta_j}^q+ X_i^T {\Sigma_{ij}^q}^{-1}(y_i-V_i     \mu_{b_j}^q)\right\}$  
\item $\Sigma_{\beta_j}^q \leftarrow \left(\Sigma_{\beta_j}^{-1}+ \frac{\alpha_{a_j}^q}{\lambda_{a_j}^q} \sum_{i=1}^n q_{ij} I_p\right)^{-1}.$
\item $\mu_{\beta_j}^q \leftarrow \Sigma_{\beta j}^q \frac{\alpha_{a_j}^q}{\lambda_{a_j}^q} \sum_{i=1}^n q_{ij} \mu_{\eta_i}^q.$ 
\stepcounter{enumi}
\item $\mu_{b_j}^q \leftarrow \Sigma_{b_j}^q \sum_{i=1}^n q_{ij} V_i^T {\Sigma_{ij}^q}^{-1} \left(y_i-X_i 
      \mu_{\eta_i}^q\right)$. 
\stepcounter{enumi}
\item $\lambda_{a_j}^q \leftarrow \lambda_{a_j} + \frac{1}{2} \sum_{i=1}^n q_{ij} 
     \{ {(\mu_{\eta_i}^q-\mu_{\beta_j}^q) }^T(\mu_{\eta_i}^q-\mu_{\beta_j}^q) 
     +\mbox{tr} (\Sigma_{\eta_i}^q + \Sigma_{\beta_j}^q) \}$. 
\stepcounter{enumi}
\stepcounter{enumi}
\stepcounter{enumi}
\item $\lambda_{jl}^q \leftarrow \lambda_{jl} + \frac{1}{2} \sum_{i=1}^n q_{ij} \left\{
(\omega_{ij})_{\kappa_{il}}^T (\omega_{ij})_{\kappa_{il}} + \mbox{tr} ( X_i\Sigma_{\eta_i}^qX_i^T 
+V_i \Sigma_{b_j}^q V_i^T)_{\kappa_{il}} \right\}$, where $\omega_{ij}=y_i-X_i\mu_{\eta_i}^q-V_i \mu_{b_j}^q$.
\stepcounter{enumi}
\item $q_{ij} \leftarrow \frac{p_{ij} \exp(c_{ij})}{\sum_{l=1}^k p_{il}\exp(c_{il})},$ where
\begin{align*}
c_{ij}= &\; \frac{p}{2} (\psi(\alpha_{a_j}^q)-\log\lambda_{a_j}^q) 
          - \frac{1}{2} \left\{ \omega_{ij}^T {\Sigma_{ij}^q}^{-1}\omega_{ij} + \mbox{tr}(
          {\Sigma_{ij}^q}^{-1} ( X_i\Sigma_{\beta_j}^q X_i^T
          +V_i \Sigma_{b_j}^q V_i^T)) \right\}        \\
        & + \sum_{l=1}^g \frac{\kappa_{il}}{2} (\psi(\alpha_{jl}^q)-\log \lambda_{jl}^q)      
          -\frac{\alpha_{a_j}^q}{2\lambda_{a_j}^q}\left\{ {(\mu_{\eta_i}^q-\mu_{\beta_j}^q) }^T(\mu_{\eta_i}^q-\mu_{\beta_j}^q) 
        +\mbox{tr} (\Sigma_{\eta_i}^q + \Sigma_{\beta_j}^q)\right\}   
\end{align*}
\end{enumerate}
The variational lower bound is given by
\begin{align*}
& \frac{1}{2}\sum_{j=1}^k \left\{\log |\Sigma_{\beta_j}^{-1} \Sigma_{\beta_j}^q|  
  - \mbox{tr} (\Sigma_{\beta_j}^{-1} \Sigma_{\beta_j}^q)
  - {\mu_{\beta_j}^q}^T \Sigma_{\beta_j}^{-1} \mu_{\beta_j}^q +\log |\Sigma_{b_j}^q| 
  - \frac{\alpha_{b_j}^q}{\lambda_{b_j}^q} 
    \left( {\mu_{b_j}^q}^T \mu_{b_j}^q + \mbox{tr} (\Sigma_{b_j}^q) \right)\right\} \\
& + \sum_{j=1}^k \left\{ \alpha_{a_j} \log \frac{\lambda_{a_j}}{\lambda_{a_j}^q}
  + \log \frac{\Gamma(\alpha_{a_j}^q)}{\Gamma(\alpha_{a_j})}
  + \frac{p}{2}\sum_{i=1}^n q_{ij} (\psi(\alpha_{a_j}^q) -\log\lambda_{a_j}^q) 
  +\psi(\alpha_{a_j}^q) (\alpha_{a_j}-\alpha_{a_j}^q)+\alpha_{a_j}^q  \right. \\
& \left. - \frac{\lambda_{a_j} \alpha_{a_j}^q}{\lambda_{a_j}^q} 
  + \alpha_{b_j} \log \frac{\lambda_{b_j}}{\lambda_{b_j}^q}
  + \log \frac{\Gamma(\alpha_{b_j}^q)}{\Gamma(\alpha_{b_j})}
  - \frac{\lambda_{b_j} \alpha_{b_j}^q}{\lambda_{b_j}^q} 
  - \frac{s_2}{2} \log\lambda_{b_j}^q + \alpha_{b_j}^q 
  + \sum_{l=1}^g \left[ \alpha_{jl} \log \frac{\lambda_{jl}}{\lambda_{jl}^q}
  + \alpha_{jl}^q \right. \right. \\
& \left. \left. + \log \frac{\Gamma(\alpha_{jl}^q)}{\Gamma(\alpha_{jl})} 
  -\sum_{i=1}^n \frac{\kappa_{il}q_{ij}}{2} \left(\psi(\alpha_{jl}^q)
  -\log\lambda_{jl}^q \right)+\psi(\alpha_{jl}^q) (\alpha_{jl}-\alpha_{jl}^q)
  - \frac{\lambda_{jl} \alpha_{jl}^q}{\lambda_{jl}^q} \right] \right\} - \frac{N}{2} \log (2\pi)\\
& -\sum_{i=1}^n \sum_{j=1}^k \frac{q_{ij}}{2} \left\{ \omega_{ij}^T
   {\Sigma_{ij}^q}^{-1} \omega_{ij} +\mbox{tr}( {\Sigma_{ij}^q}^{-1} ( X_i\Sigma_{\eta_i}^qX_i^T+V_i  
   \Sigma_{b_j}^q V_i^T))-\frac{\alpha_{a_j}^q}{\lambda_{a_j}^q} \left(\mbox{tr} (\Sigma_{\eta_i}^q+\Sigma_{\beta_j}^q)+ \right. \right. \\
& \left. \left.(\mu_{\eta_i}^q-\mu_{\beta_j}^q)^T (\mu_{\eta_i}^q-\mu_{\beta_j}^q) \right)
 -2\log \frac{p_{ij}}{q_{ij}} \right\}
 +\frac{k(p+s_2)+np}{2}+ \frac{1}{2}\sum_{i=1}^n \log |\Sigma_{\eta_i}^q|+\log p(\mu_\delta^q)
\end{align*}
where $\omega_{ij}=y_i-X_i\mu_{\eta_i}^q-V_i \mu_{b_j}^q$. 
In the examples, when Algorithm 2 is used in conjunction 
with the VGA to fit a 1-component mixture ($j=1$), we set
$q_{ij}=1$ for $i=1,...,n$, $\frac{\alpha_{jl}^q}{\lambda_{jl}^q}=1$ for $l=1,...,g$,
$\frac{\alpha_{b_j}^q}{\lambda_{b_j}^q}=1$, $\frac{\alpha_{a_j}^q}{\lambda_{a_j}^q}=0.1$, 
$\mu_{b_j}^q=0$ and $\mu_{\beta_j}^q=0$ for initialization . 

\section*{Algorithm 3 (full centering when $X_i=W_i=V_i$)}
(Updates in steps 7, 9, 11 and 13 remain the same as in algorithm 1 with $s_1=s_2=p$.)
Initialize: $q_{ij}$ for $i=1,...,n$, $j=1,...,k$, $\mu_{\nu_j}^q$, $\mu_{\beta_j}^q$ 
$\frac{\alpha_{a_j}^q}{\lambda_{a_j}^q}$ and $\frac{\alpha_{b_j}^q}{\lambda_{b_j}^q}$
for $j=1,...,k$, $\frac{\alpha_{jl}^q}{\lambda_{jl}^q}$ for $j=1,...,k$, $l=1,...,g$.
Do until the change in the lower bound between iterations is less than a tolerance:
\begin{enumerate}
\item $\Sigma_{\rho_i}^q \leftarrow \left( \sum_{j=1}^k q_{ij} \frac{\alpha_{a_j}^q}{\lambda_{a_j}^q} I_p 
		  + X_i^T (\sum_{j=1}^k q_{ij} {\Sigma_{ij}^q}^{-1}) X_i \right)^{-1}.$  
\item $\mu_{\rho_i}^q \leftarrow \Sigma_{\rho_i}^q \sum_{j=1}^k q_{ij} 
      \left( \frac{\alpha_{a_j}^q}{\lambda_{a_j}^q} \mu_{\nu_j}^q+ X_i^T {\Sigma_{ij}^q}^{-1} y_i \right).$
\item $\Sigma_{\nu_j}^q \leftarrow \left(\left(\frac{\alpha_{b_j}^q}{\lambda_{b_j}^q}
			+\frac{\alpha_{a_j}^q}{\lambda_{a_j}^q}\sum_{i=1}^n q_{ij}\right)I_p \right)^{-1}.$  
\item $\mu_{\nu_j}^q \leftarrow \Sigma_{\nu_j}^q  
      \left( \frac{\alpha_{b_j}^q}{\lambda_{b_j}^q} \mu_{\beta_j}^q
      + \frac{\alpha_{a_j}^q}{\lambda_{a_j}^q} \sum_{i=1}^n q_{ij}\mu_{\rho_i}^q \right).$
\item $\Sigma_{\beta_j}^q \leftarrow \left(\Sigma_{\beta_j}^{-1}+
       \frac{\alpha_{b_j}^q}{\lambda_{b_j}^q} I_p \right)^{-1}.$
\item $\mu_{\beta_j}^q \leftarrow \Sigma_{\beta j}^q \frac{\alpha_{b_j}^q}{\lambda_{b_j}^q}\mu_{\nu_j}^q.$
\stepcounter{enumi}
\item $\lambda_{a_j}^q \leftarrow \lambda_{a_j} + \frac{1}{2} \sum_{i=1}^n q_{ij} 
     \{ {(\mu_{\rho_i}^q-\mu_{\nu_j}^q) }^T(\mu_{\rho_i}^q-\mu_{\nu_j}^q) 
     +\mbox{tr} (\Sigma_{\rho_i}^q + \Sigma_{\nu_j}^q) \}.$
\stepcounter{enumi}
\item $\lambda_{b_j}^q \leftarrow \lambda_{b_j} + \frac{1}{2}  
     \{ {(\mu_{\nu_j}^q-\mu_{\beta_j}^q) }^T(\mu_{\nu_j}^q-\mu_{\beta_j}^q) 
     +\mbox{tr} (\Sigma_{\nu_j}^q + \Sigma_{\beta_j}^q)\}.$
\stepcounter{enumi}
\item $\lambda_{jl}^q \leftarrow \lambda_{jl} + \frac{1}{2} \sum_{i=1}^n q_{ij} \left\{
           (y_i-X_i\mu_{\rho_i}^q)_{\kappa_{il}}^T (y_i-X_i\mu_{\rho_i}^q)_{\kappa_{il}}  
           + \mbox{tr} ( X_i\Sigma_{\rho_i}^qX_i^T)_{\kappa_{il}} \right\}.$  
\stepcounter{enumi}
\item $q_{ij} \leftarrow \frac{p_{ij} \exp(c_{ij})}{\sum_{l=1}^k p_{il}\exp(c_{il})},$ where
\begin{align*}
c_{ij}= & \frac{p}{2} \left( \psi(\alpha_{a_j}^q)-\log\lambda_{a_j}^q\right)
  - \frac{1}{2}\left( (y_i-X_i\mu_{\rho_i}^q)^T {\Sigma_{ij}^q}^{-1}(y_i-X_i\mu_{\rho_i}^q)
  + \mbox{tr}({\Sigma_{ij}^q}^{-1}X_i\Sigma_{\rho_i}^q X_i^T) \right)\\
& -\frac{\alpha_{a_j}^q}{2\lambda_{a_j}^q}
   \left( {(\mu_{\rho_i}^q-\mu_{\nu_j}^q) }^T(\mu_{\rho_i}^q-\mu_{\nu_j}^q) 
  +\mbox{tr} (\Sigma_{\rho_i}^q + \Sigma_{\nu_j}^q)\right) 
  + \sum_{l=1}^g \frac{\kappa_{il}}{2} \left( \psi(\alpha_{jl}^q)-\log\lambda_{jl}^q\right). 
\end{align*} 
\end{enumerate}
The variational lower bound is given by
\begin{align*}
& -\frac{1}{2}\sum_{j=1}^k \left\{\log |\Sigma_{\beta_j}^{-1} \Sigma_{\beta_j}^q| 
  +\mbox{tr}(\Sigma_{\beta_j}^{-1} \Sigma_{\beta_j}^q)
  +{\mu_{\beta_j}^q}^T \Sigma_{\beta_j}^{-1} \mu_{\beta_j}^q -\log|\Sigma_{\nu_j}^q| \right\}
  + \frac{1}{2}\sum_{i=1}^n \log |\Sigma_{\rho_i}^q|\\ 
& +\sum_{j=1}^k \left\{ \alpha_{a_j} \log \frac{\lambda_{a_j}}{\lambda_{a_j}^q}
  + \log \frac{\Gamma(\alpha_{a_j}^q)}{\Gamma(\alpha_{a_j})}
  +\alpha_{a_j}^q + \frac{p}{2}\sum_{i=1}^n q_{ij}(\psi(\alpha_{a_j}^q)-\log\lambda_{a_j}^q)
  + \psi(\alpha_{a_j}^q)(\alpha_{a_j}-\alpha_{a_j}^q) \right. \\
& \left. - \frac{\lambda_{a_j} \alpha_{a_j}^q}{\lambda_{a_j}^q} 
  +\alpha_{b_j} \log \frac{\lambda_{b_j}}{\lambda_{b_j}^q} 
  + \log \frac{\Gamma(\alpha_{b_j}^q)}{\Gamma(\alpha_{b_j})}
  - \frac{\lambda_{b_j} \alpha_{b_j}^q}{\lambda_{b_j}^q}
  - \frac{p}{2}\log\lambda_{b_j}^q+ \alpha_{b_j}^q \right\} 
  + \sum_{j=1}^k \sum_{l=1}^g \left\{ \alpha_{jl} \log \frac{\lambda_{jl}}{\lambda_{jl}^q} \right.    \\  
& \left. + \psi(\alpha_{jl}^q)(\alpha_{jl}-\alpha_{jl}^q)
  + \log \frac{\Gamma(\alpha_{jl}^q)}{\Gamma(\alpha_{jl})} 
  +  \sum_{i=1}^n \frac{\kappa_{il}q_{ij}}{2} \left(\psi(\alpha_{jl}^q)- \log\lambda_{jl}^q\right) 
  - \frac{\lambda_{jl} \alpha_{jl}^q}{\lambda_{jl}^q} + \alpha_{jl}^q  \right\}  \\
& -\sum_{j=1}^k \frac{\alpha_{b_j}^q}{2\lambda_{b_j}^q} 
   \left((\mu_{\nu_j}^q-\mu_{\beta_j}^q)^T (\mu_{\nu_j}^q-\mu_{\beta_j}^q)
  +\mbox{tr} (\Sigma_{\nu_j}^q+\Sigma_{\beta_j}^q)\right)
  +\sum_{i=1}^n \sum_{j=1}^k q_{ij}\log \frac{p_{ij}}{q_{ij}} +\frac{p(2k+n)}{2}  \\
& -\sum_{i=1}^n \sum_{j=1}^k \frac{q_{ij}}{2} 
    \left\{ (y_i-X_i\mu_{\rho_i}^q)^T {\Sigma_{ij}^q}^{-1} (y_i-X_i\mu_{\rho_i}^q)
  + \mbox{tr} ({\Sigma_{ij}^q}^{-1}( X_i\Sigma_{\rho_i}^qX_i^T))  \right\}+ \log p(\mu_\delta^q)\\
& -\sum_{i=1}^n \sum_{j=1}^k \frac{q_{ij}\alpha_{a_j}^q}{2\lambda_{a_j}^q} \left(\mbox{tr} (\Sigma_{\rho_i}^q+\Sigma_{\nu_j}^q)+(\mu_{\rho_i}^q-\mu_{\nu_j}^q)^T (\mu_{\rho_i}^q-\mu_{\nu_j}^q) \right)- \frac{N}{2} \log (2\pi).
\end{align*}
In the examples, when Algorithm 3 is used in conjunction 
with the VGA to fit a 1-component mixture ($j=1$), we set 
$q_{ij}=1$ for $i=1,...,n$, 
$\frac{\alpha_{jl}^q}{\lambda_{jl}^q}=10$ for $l=1,...,g$, $\frac{\alpha_{a_j}^q}{\lambda_{a_j}^q}=0.1$,
$\frac{\alpha_{b_j}^q}{\lambda_{b_j}^q}=0.01$, $\mu_{\beta_j}^q=0$ 
and $\mu_{\nu_j}^q=0$ for initialization.
We note that the rate of convergence of 
Algorithm 3 can be sensitive to the initialization of $\frac{\alpha_{jl}^q}{\lambda_{jl}^q}$, $\frac{\alpha_{a_j}^q}{\lambda_{a_j}^q}$ and $\frac{\alpha_{b_j}^q}{\lambda_{b_j}^q}$ and observed 
that an initialization satisfying
$\frac{\alpha_{b_j}^q}{\lambda_{b_j}^q}<\frac{\alpha_{a_j}^q}{\lambda_{a_j}^q} <\frac{\alpha_{jl}^q}{\lambda_{jl}^q} $ works better.

\section*{Example on clustering of yeast galactose data}
The yeast galactose data of Ideker {\it et al.} (2001) has four replicate 
hybridizations for each of 20 cDNA array experiments. We consider a subset 
of 205 genes previously analyzed by Yeung {\it et al.} (2003) and Ng {\it et al.}
(2006) whose expression patterns reflect four functional categories in the 
Gene Ontology (GO) listings (Ashburner {\it et al.}, 2000). This dataset is available from the online version of Yeung {\it et al.} (2003). Approximately 8\% of the data are missing and Yeung {\it et al.} (2003)
used a $k$-nearest neighbour method to impute the missing data values. 
Yeung {\it et al.} (2003) and Ng {\it et al.} (2006) evaluated the performance of their clustering algorithms 
by how closely the clusters compared with the four categories in the GO 
listings. They used the adjusted Rand index (Hubert and Arabie, 1985) 
to assess the degree of agreement between their partitions and the four 
functional categories.

We use this example to illustrate the way that our model can make use of 
covariates in the mixing weights, unlike previous analyses of this data set.
In particular, we use the GO listings as covariates in the mixture weights. Let $u_i$ be a vector of length $d=4$ where the $l$th element is 1 if the functional category of gene $i$ is 
$l$ and 0 otherwise. Instead of looking at the data 
with the missing values imputed by the $k$-nearest neighbour method, 
we consider the original data containing 8\% missing values, 
since our model has the capability to handle missing data. Taking $n=205$ 
genes, let $y_{itr}$ denote the $r$th repetition of the expression profile for 
gene $i$ at experiment $t$, $0 \leq r \leq 4$, and $R_{it}$ denote the 
number of replicate hybridizations data available for gene $i$ in experiment
$t$, $i=1,...,205$, $t=1,...,20$. For each $i=1,...,n$, $y_i$ is a vector 
of $n_i$ observations where $n_i=\sum_{t=1}^{20}R_{it}$ and 
$y_i = (y_{i11},...,y_{i14},...,y_{i,20,1},...,y_{i,20,4})^T,$
with missing observations omitted. 
$V_i$ is a $n_i \times 80$ matrix obtained from 
$I_{80}$ by removing the $(tr)$th row if the observation for 
experiment $t$ at the $r$th repetition is not available.
$X_i$ is a $n_i \times 20$ matrix, 
 $$X_i=\left( \begin{matrix} 
1_{R_{i1}}   & 0_{R_{i1}}  & ...    & 0_{R_{i1}} \\
0_{R_{i2}}   & 1_{R_{i2}}  & ...    & 0_{R_{i2}} \\
 \vdots      & \vdots      & \ddots & \vdots  \\
0_{R_{i20}}  & 0_{R_{i20}} & ...    & 1_{R_{i20}} \end{matrix} \right),$$
and $W_i=X_i$. For the error 
terms, we set $g=20$ with $\kappa_{il}=R_{il}$, $i=1,...,n$, $l=1,...,g$, so that 
the error variance of each mixture component is allowed to vary between different experiments. 
We used the following priors, $\delta\sim N(0,1000I)$,
$\beta_j\sim N(0,1000I)$ for $j=1,...,k,$ and $IG(0.01,0.01)$ 
for $\sigma_{a_j}^2$, $\sigma_{b_j}^2$, $j=1,...,k$ and $\sigma_{jl}^2$,
$j=1,...,k$, $l=1,...,g$.

\begin{figure}
\begin{center}
\includegraphics[width=150mm]{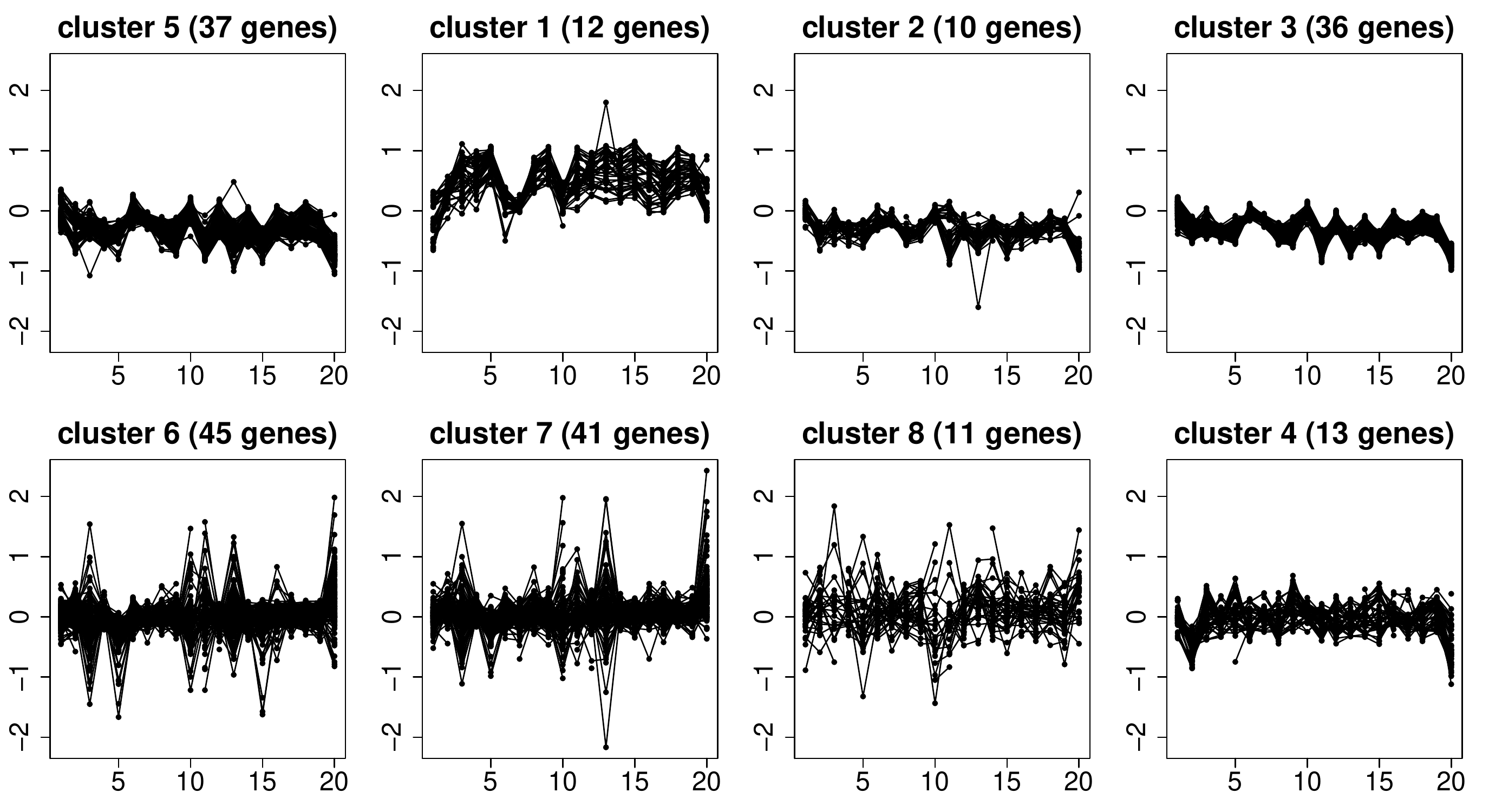} 
\end{center}
\caption{\label{yeast} Clustering results for yeast galactose data from applying the VGA using Algorithm 2. The $x$-axis are the experiments and $y$-axis are the gene expression profiles. GO listings were used as covariates in the mixture weights. }
\end{figure}

\begin{figure}
\begin{center}
\includegraphics[width=90mm]{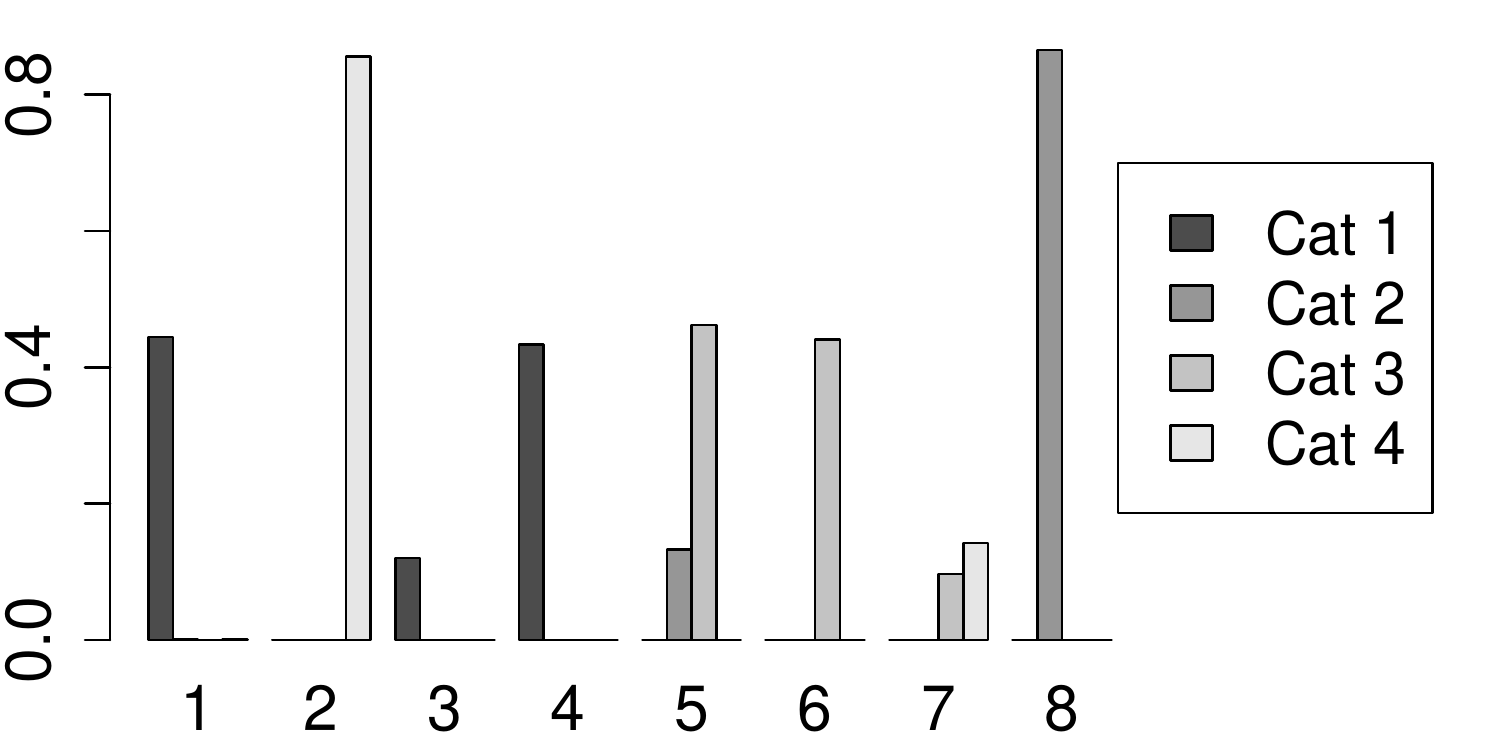} 
\end{center}
\caption{\label{postprob} Fitted probabilities from gating function. The $x$-axis are the clusters
and $y$-axis are the probabilities.}
\end{figure}

Applying the VGA using Algorithm 2 five times, we obtained a 8-component mixture on all five trials with  similar results. The clustering of a 8-component mixture with the highest estimated log marginal likelihood among the five trials is shown in Figure \ref{yeast}. 
Our clustering results are slightly different from Ng {\it et al} (2006) where the number of optimal clusters obtained was 7. We note that genes from Category 1 were split into 3 clusters by the
VGA on all 5 attempts while in Ng {\it et al} (2006), genes from Category 1 were subdivided 
into 2 clusters. In addition, instead of having one cluster containing all the genes from 
Category 4, we observed that two or three of the genes in Category 4 were consistently separated 
from the cluster containing the remaining genes from Category 4. Fitted probabilities from the gating function are shown in Figure \ref{postprob}. 
These were obtained by substituting $\delta$ with $\mu_\delta^q$ from the variational posterior into 
$P(z_i=j)=p_{ij}=\frac{\exp(u_i^T \delta_j)}{\sum_l \exp(u_i^T \delta_l)}$
which represents the probability that observation $i$ belongs to component $j$ of the mixture conditional on the category that observation $i$ belongs to in the GO listings. 

To investigate the impact of reparametrizing the model using hierarchical centering, we applied VGA using Algorithm 1 five times. This time, we obtained a 8-component mixture four times and a 9-component mixture  once. The average estimated log marginal log likelihood attained by Algorithm 1 was 8491, lower than the average of 8999 attained by Algorithm 2. For fitting a 8-component model, VGA with Algorithm 1 took an average of 8455 seconds, while Algorithm 2 took an average of 3185 seconds. While these results may not be conclusive, the gain in efficiency in using Algorithm 2 over Algorithm 1 is clear. By using hierarchical centering, the computation time was reduced by more than half in this example.

\section*{References}
\bib
Ashburner, M., Ball, C.A., Blake, J.A., Botstein, D., Butler, H., Cherry, J.M., Davis, A.P., Dolinski, K., 
Dwight, S.S., Eppig, J.T., Harris, M.A., Hill, D.P., Issel-Tarver, L., Kasarskis, A., Lewis, S., Matese, J.C., 
Richardson, J.E., Ringwald, M., Rubin, G.M. and Sherlock, G. (2000).
Gene ontology: tool for the unification of biology.
{\it Nature Genetics}, 25, 25--29.

\bib
Hubert, L. and Arabie, P. (1985).
Comparing partitions.
{\it Journal of Classification}, 2, 193--218.

\bib
Ideker, T., Thorsson, V., Ranish, J.A., Christmas, R., Buhler, J., Eng, J.K., Bumgarner, R., 
Goodlett, D.R., Aebersold, R. and Hood, L. (2001). 
Integrated genomic and proteomic analyses of a systematically perturbed metabolic network.
{\it Science}, 292, 929--934.  

\bib
Ng, S.K., McLachlan, G.J., Wang, K., Ben-Tovim Jones, L. and Ng, S.-W. (2006).  
A mixture model with random-effects components for clustering correlated gene-expression profiles.
{\it Bioinformatics}, 22, 1745--1752. 

\bib
Yeung, K.Y., Medvedovic, M. and Bumgarner, R.E. (2003).
Clustering gene-expression data with repeated measurements.
{\it Genome Biology}, 4, Article R34.